\documentclass[twocolumn,showpacs,preprintnumbers,amsmath,amssymb]{revtex4}
\usepackage{epsfig}% Include figure files
\usepackage{psfig}% Include figure files
\usepackage{graphicx}% Include figure files
\usepackage{dcolumn}% Align table columns on decimal point
\usepackage{bm}% bold math
\usepackage{subfigure}

\newcommand{\be}{\begin{equation}}    % for lazy typers
\newcommand{\ee}{\end{equation}}
\newcommand{\beq}{\begin{eqnarray}}
\newcommand{\eeq}{\end{eqnarray}}
\newcommand{\beqn}{\begin{eqnarray*}}
\newcommand{\eeqn}{\end{eqnarray*}}
\def\lsim{\mathrel{\rlap{\lower2.5pt\hbox{\hskip1pt$\sim$}}
    \raise1pt\hbox{$<$}}}         %less than or approx. symbol
\def\gsim{\mathrel{\rlap{\lower2.5pt\hbox{\hskip1pt$\sim$}}
    \raise1pt\hbox{$>$}}}         %greater than or approx. symbol
\def\msun{M_\odot}
\def\nn{\nonumber}
\def\apr2{{\rm APR2~}}
\def\s1{{\rm BBS1~}}
\def\bs2{{\rm BBS2~}}
\def\g240{{\rm G240~}}
\def\ss1{{\rm SS1~}}
\def\sc2{{\rm SS2~}}
\def\lappreq{\mathrel{\rlap{\lower2.5pt\hbox{\hskip1pt$\sim$}}
    \raise1pt\hbox{$<$}}}         %less than or approx. symbol
\def\gappreq{\mathrel{\rlap{\lower2.5pt\hbox{\hskip1pt$\sim$}}
    \raise1pt\hbox{$>$}}}         %greater than or approx. symbol

\begin{document}

%%%%%%%%%%%%%%%%%%%%%%%%%%%%%%%%%%%%%%%%%%%%%%%%

%%%%%%%%%%%%%%%%%%%%TTTTT%%%%%%%%%%%%%%%%%%%%%%%%%%%%%%%%%%%%%%%%

\title{Perturbative approach to the structure of rapidly rotating neutron stars}
\author{Omar Benhar$^{1,2}$, Valeria Ferrari$^{1,2}$,  Leonardo Gualtieri$^{1,2}$,
Stefania Marassi$^{1}$}
\affiliation{ $^1$  Dipartimento di Fisica ``G. Marconi",
Universit\'a degli Studi di Roma, ``La Sapienza", P.le A. Moro
2, 00185 Roma, Italy\\
$^2$ INFN, Sezione Roma 1,P.le A. Moro
2, 00185 Roma, Italy
}

\date{\today}

\label{firstpage}

\begin{abstract}
We construct models of rotating stars using the perturbative approach
introduced by J. Hartle in 1967, and a set of equations of state
proposed to model hadronic interactions in the inner core of
neutron stars.  We integrate the equations of stellar structure to
third order in the angular velocity and show, comparing our results to
those obtained with fully non linear codes, to what extent 
third order corrections are needed to accurately reproduce the moment of 
inertia of a star which rotates at rates comparable to that
of the  fastest isolated pulsars.
\end{abstract}

\maketitle
%%%%%%%%%%%%%%%%%%%%%%%%%%%%%%%%%%%%%%%%%%%%%%%%%%%%%%%%%%%%%%%%%%%%%%
\section{Introduction}
%%%%%%%%%%%%%%%%%%%%%%%%%%%%%%%%%%%%%%%%%%%%%%%%%%%%%%%%%%%%%%%%%%%%%%
In this paper we show that the structure of a rapidly rotating neutron
star can be described to a high level of accuracy by using the
perturbative approach introduced by Hartle in 1967 \cite{hartle},
subsequently developed to third order in the angular velocity $\Omega$
\cite{hartle1}.

There are several reasons why the perturbative approach may be
preferred to the direct numerical integration of 
Einstein's equations for stellar structure.  One is that the angular
behaviour of the solution is given explicitely in terms of Legendre
polynomials of first kind, whereas the radial behaviour can be found
by integrating linear differential equations in the radial variable;
thus, instead of solving the nonlinear, coupled, partial differential
equations of the exact theory one simply integrates a set of ordinary,
linear differential equations. Moreover, these equations need to be
integrated only inside the star, because the exterior solution can be
found explicitely in terms of known polynomials.  A further element of
interest is that the perturbative approach allows to evaluate to what
extent a physical quantity changes at each order in $\Omega$, and to
check whether the $n^{th}$-order contribution is needed to estimate a
choosen quantity or whether the first order estimates, often used to
interpret astronomical observations, are enough. 
%LLLLLLLLLLLLLLLLLLLLLLLLLLLLLLLLLLLLLLLLLLLLLLLLLLLLLLLLLL
For these reasons, the perturbative approach expanded to 2nd order in
rotation rate has been used by several authors to study properties of rotating
neutron stars \cite{hartlet,chanmil,weberglen1,weberglen2,weberglen3,latpra,marco} 
(see also the review \cite{sterg} and refereces therein).
%LLLLLLLLLLLLLLLLLLLLLLLLLLLLLLLLLLLLLLLLLLLLLLLLLLLLLLLLLL

On the other hand the perturbative approach has a drawback, since it
fails when the angular velocity approaches the mass-shedding limit,
i.e. when the star rotates so fastly that the gravitational attraction
is not sufficient to keep matter bound to the surface; numerical
integration of the exact equations shows that in these conditions a
cusp forms on the equatorial plane of the star
\cite{bonazzola,bonazzola1,cook2}, which cannot be reproduced by the
first few orders of the perturbative expansions (we shall discuss this
point quantitatively in section \ref{section:sec3}).  However, the
rotation rate of all known pulsars is much lower than the
mass-shedding limit for most equations of state (EOS) considered in the
literature. Thus, unless we are specifically interested in the
mass-shedding problem, we can use the perturbative approach to study
the structural properties of the observed neutron stars.  The
problem is to establish how many terms in the expansion we need to consider to
describe, say, the fastest isolated pulsar observed so far, PSR
1937+21, whose period is $T= 1.56$ ms, and this is the question we
plan to answer in this paper.
The equations of structure for uniformly rotating stars have been
developed to second order in the angular velocity in
\cite{hartle,hartlet} and subsequently extended to third order in
\cite{hartle1}.  With respect to the derivation of \cite{hartle1} we
introduce two novelties:\\
%---------------------------------------------------------------------------------
- the third order corrections involve two functions, $w_1$ and $w_3.$
$w_1$ gives a contribution to the moment of inertia, $w_3$ affects the
mass-shedding velocity. Both $w_1$ and $w_3,$ contribute to the
dragging of inertial frames.  In \cite{hartle1} the solution for $w_1$
was given explicitely outside the star in terms of Legendre
polynomials of second kind; that expression contained some errors that
we correct.  In addition, we complete the third order solution giving
the explicit expression of $w_3$ outside the star.\\
%---------------------------------------------------------------------------------
- In \cite{hartle1} a procedure was developed to construct families of
constant baryonic mass solutions with varying angular velocity, which
was applied to study the behaviour of the moment of inertia along such
sequencies, for small values of the angular velocities.  This
procedure cannot be applied to fastly rotating stars and furthermore,
%because the correction to the central energy density is treated as a
%perturbation, and this correction becomes large while approaching the
%mass shedding limit. 
it was shown to be unstable when the stellar mass approaches the
maximum mass for any assigned equation of state.  In this paper we
develop, and apply, a different algorithm that allows to describe a
sequence of constant baryonic mass stars at any value of the angular
velocity (smaller than the mass-shedding limit), and which is stable
also in the maximum mass limit.
%---------------------------------------------------------------------------------

In section \ref{section:sec2} we shall briefly explain what are the
corrections to the metric and to the thermodynamical functions that
are needed to describe the structure of a rotating star to third order
in $\Omega$, and in the appendix we shall summarize the equations they
satisfy inside and outside the star.

In order to test the perturbative scheme, we shall compare the results
we find solving these equations for three selected EOS, with those of
ref.  \cite{cook2} and \cite{EN} where the exact Einstein equations have
been integrated for the same EOS.  
%LLLLLLLLLLLLLLLLLLLLLLLLLLLLLLLLLLLLLLLLLLLLLLLLLLLLLLLLLLLLLL
To hereafter, the adjective ``exact" referred to Einstein's equations
will mean that the full set of equations is solved numerically and 
non-perturbatively.
%LLLLLLLLLLLLLLLLLLLLLLLLLLLLLLLLLLLLLLLLLLLLLLLLLLLLLLLLLLLLLL
This comparison will be discussed
in section \ref{section:sec3} and will allow us to assess the validity
of the third order approach.

We shall then integrate the third order equations to model a neutron
star composed of an outer crust, an inner crust and a core, each shell
being described by a non viscous fluid which obeys equations of state
appropriate to describe different density regions.  There is a general
consensus on the matter composition of the crust, for which we use two
well established EOS: the Baym-Pethick-Sutherland (BPS) EOS \cite{BPS}
for the outer crust ($ 10^7 \lappreq \rho \lappreq 4\cdot 10^{11}$
g/cm$^3$), and the Pethick-Ravenhall-Lorenz (PRL) EOS \cite{PRL} for
the inner crust ($4\cdot 10^{11} \lappreq \rho \lappreq 2\cdot
10^{14}$ g/cm$^3$).

For the inner core, $\rho > 2\cdot 10^{14}$ g/cm$^3$, we use recent
EOS which model hadronic interactions in different ways leading to
different composition and dynamics. The main assumptions underlying
these EOS will be outlined in section \ref{section:eos}.

For a given EOS there are two kind of equilibrium configurations,
ususally indicated as "normal" and "supramassive". They both refer to
rotating stars with constant baryonic mass and varying angular
velocity. Each normal sequence has, as a limiting configuration, a non
rotating, spherical star. The supramassive sequences do not possess
this limit, and all members of a sequence have a baryonic mass larger
than the maximum mass of a non rotating, spherical star.  Our study is
confined to stars belonging to the normal sequence.

The results of our calculations will be discussed in section
\ref{section:results}.

%%%%%%%%%%%%%%%%%%%%%%%%%%%%%%%%%%%%%%%%%%%%%%%%%%%%%%%%%%%%%%%%%%%%%%%%%%%%%
\section{The metric and the structure of a rotating star} \label{section:sec2}
%%%%%%%%%%%%%%%%%%%%%%%%%%%%%%%%%%%%%%%%%%%%%%%%%%%%%%%%%%%%%%%%%%%%%%%%%%%%%
In this section we introduce all the quantities that describe the star
to third order in the angular velocity $\Omega$.  Following
\cite{hartle,hartle1,hartlet} we write the metric as
\beq
\label{metric1}
ds^2&=&-e^{\nu(r)}\left[1+2h(r,\theta)\right] dt^2\\
&+& e^{\lambda(r)} \left[1+\frac{2m(r,\theta)}{[r-2M(r)]}\right] 
dr^2\nonumber\\
&+&r^2\left[1+2k(r,\theta)\right]
\left\{ d\theta^2+
\sin^2\theta~[d\phi-w(r,\theta) dt ]^2\right\} .
\nonumber
\eeq
Here $e^{\nu(r)}$, $e^{\lambda(r)}$ and $M(r)$ are functions of $r$
and describe the non rotating star solution of the TOV equations (see
Appendix A).  The functions $h(r,\theta)$, $m(r,\theta)$,
$k(r,\theta)$ and $w(r,\theta)$ are the perturbative corrections
\beq
\label{expansion}
h(r,\theta)&=&h_0(r)+h_2(r)P_2(\theta)+ O(\Omega^4)\\
\label{hexp}
m(r,\theta)&=&m_0(r)+m_2(r)P_2(\theta)+ O(\Omega^4)\\
\label{mexp}
k(r,\theta)&=&k_2(r)P_2(\theta)\\
\nonumber
&=&\left[ v_2(r)-h_2(r)\right]~ P_2(\theta)+ O(\Omega^4)\\
\label{kexp}
w(r,\theta)&=&\omega(r) + w_1(r)\\
\label{oexp}
\nn
&-&w_3(r) \frac{1}{\sin\theta}\frac{dP_3(\theta)}{d\theta} +
O(\Omega^5)
\eeq
$P_2(\theta)$ and $P_3(\theta)$ are the $l=2$ and $l=3$ Legendre polynomials.
As described in \cite{hartle,hartle1,hartlet},
the function $\omega$ is of order $\Omega$, whereas  $(h_0, h_2, m_0, m_2,
k_2, v_2)$ are of order $\Omega^2$, and $(w_1,  w_3)$ are of order $\Omega^3$.
In a similar way, the displacement that an element of fluid, located 
at a given $(r,\theta)$ in the non rotating star, experiences because of rotation 
can be expanded in even powers of $\Omega$
\be
\label{displ1}
\xi=\xi_0(r)+\xi_2(r)P_2(\theta)+ O(\Omega^4),
\ee
where both $\xi_0(r)$ and $\xi_2(r)$ are of order $\Omega^2$.
We shall assume that matter in the star is described by a perfect fluid with  energy momentum
tensor
\be
\label{enmom}
T^{\mu\nu} = (\epsilon +P) u^\mu u^\nu + P g^{\mu\nu}.
\ee
Since the fluid element is displaced, its energy density
and pressure change; in a reference
frame that is momentarily moving with the fluid, 
the pressure variation  is
\be
\label{pres1}
\delta P (r,\theta) =
\left[ \epsilon(r)+P(r)\right] \left[\delta p_0(r)+\delta p_2(r)P_2(\theta)\right]\,
\ee
and it can be expressed in terms of $\xi$ as follows
\beq
\label{pres0}
\delta p_0(r)&=&-\xi_0(r)\left[\frac{1}{\epsilon+P}\frac{dP}{dr}\right],\\
\delta p_2(r)&=&-\xi_2(r)\left[\frac{1}{\epsilon+P}\frac{dP}{dr}\right],
\label{pres2}
\eeq
where $\epsilon(r)$ and $P(r)$ are the energy density and the pressure
computed for the non rotating configuration. The corresponding energy
density variation of the fluid element is
\be
\label{en1}
\delta\epsilon (r,\theta) =\frac{d\epsilon}{dP}
(\epsilon+P)\left[ \delta p_0(r)+\delta p_2(r)P_2(\theta)\right]\,.
\ee
This last equation has been derived on  the assumption that the matter satisfies a
barotropic EOS.

In appendix A we shall write the equations satisfied by the
corrections to the metric functions and to the thermodynamical
variables to the various order in $\Omega$, with the appropriate
boundary conditions.  It should be stressed that these equations have
to be numerically integrated only for $r \le R$, where $R$ is the
radius of the non rotating star. For $r > R$ the solution can be found
analytically.  Hereafter, $r=R$ and $M(R)$ will indicate,
respectively, the radius and the mass of the non rotating star, found
by solving the TOV equations for an assigned EOS.

Once the various functions have been computed as shown in Appendix A,
we can evaluate how the star changes in shape due to rotation, the
moment of inertia, the mass-shedding limit and all quantities one is
interested in.  Here we shall briefly summarize what are the relevant
physical quantities we shall consider.

\begin{itemize}
\item{\it Stellar deformation}

The effect of rotation described by the metric (\ref{metric1}) on the
shape of the star can be divided in two contributions:\\ 
A) A spherical expansion which changes the radius of the star, and is
described by the functions $h_0$ and $m_0$.\\ 
B) A quadrupole deformation, described by the functions $h_2, ~v_2$
and $m_2$.

As a consequence of these two contributions, that are both of order
$\Omega^2$ and that will be indicated with the labels A and B,
respectively, the radius of the star, the eccentricity and the
gravitational mass change as follows.

A fluid element on the surface of the spherical non rotating star, i.e. 
at $(r=R,\theta),$ will be displaced by an amount $\delta R$ given by
\be
\delta R= \delta R^A + \delta R^B= \xi_0(R) + \xi_2(R)P_2(\theta)
\label{deltaraggioA}
\ee
where $\xi_0(R)$ and $\xi_2(R)$ can be computed by
eqs. (\ref{deltaraggio}) and (\ref{xi2}) as shown in appendix A.

Once we know $\delta R$ we can compute, for instance the eccentricity of the star
\be
\label{ecc}
E_c = \left[ \hbox{(equatorial~radius)}^2 / \hbox{(polar~radius)}^2 -1\right]^{1/2}.
\ee

The change in the gravitational mass is given by
\be
\label{masschange}
\delta M_{grav}=  \left[m_0(R)+\frac{J^2}{R^3}\right].
\ee
Notice that $\delta M_{grav}$ does not depend on quadrupolar
perturbations ($h_2, v_2, m_2$), nor on the perturbations that are
first ($\omega$) and third order ($w_1, w_3$) in $\Omega$. The reason
is that such quantity is invariant under rotations of the system, and
it is even for parity transformations (which change
$\Omega\rightarrow-\Omega$).

\item{\it Dragging of inertial frames}

The function $\omega(r,\theta)$ in the metric (\ref{metric1}) is
responsible for the dragging of inertial frames. It affects two
important physical quantities, i.e.  the mass-shedding limit (section
\ref{subsection:massshedd}) and the moment of inertia (section
\ref{subsection:inerzia}).
\end{itemize}

%%%%%%%%%%%%%%%%%%%%%%%%%%%%%%%%%%%%%%%%%%%%%%%%%%%%%%%%%%%%%%%%%%%%%%%%%%%%%
\subsection{The mass-shedding limit}
\label{subsection:massshedd}
%%%%%%%%%%%%%%%%%%%%%%%%%%%%%%%%%%%%%%%%%%%%%%%%%%%%%%%%%%%%%%%%%%%%%%%%%%%%%
The mass-shedding velocity can be found using a procedure developed by
\cite{Fried}.  Given a star with assigned baryonic mass which rotates
at a given $\Omega$, let us consider an element of fluid which belongs
to the star and is located on the surface at the equator; from the
metric (\ref{metric1}) it is easy to see that it moves with velocity
\be
\label{vel}
V^{bound}=e^{\frac{\psi(r,\theta)-\beta(r,\theta)}{2}}\left[\Omega -
\left(\omega(r)+w_1(r)-\frac{3}{2}w_3(r)\right) \right],
\ee
where
\beq
\label{nupsi}
e^{\beta(r,\theta)}&=&e^{\nu(r)}\left[1+2h(r,\theta)\right] \\
\nonumber
e^{\psi(r,\theta)}&=& r^2\left[1+2k(r,\theta)\right]\sin^2\theta .
\eeq
In general, the fluid element will not follow a geodesic of metric
(\ref{metric1}).

Using the geodesic equations we can also compute the velocity of a
particle which moves on a circular orbit just outside the equator,
in the co-rotating direction
\beq
\label{veleq}
&&V^{free}=\frac{(\omega^\prime +w_1^\prime -\frac{3}{2}w_3^\prime
)}{\psi^\prime }e^{\frac{\psi -\beta}{2}}\\
\nonumber
&&+\left\{ \frac{\beta^\prime }{\psi^\prime }+
\left[\frac{(\omega^\prime +w_1^\prime -\frac{3}{2}w_3^\prime )}
{\psi^\prime }e^{\frac{\psi-\beta}{2}}
\right]^2\right\}^{1/2},
\eeq
where a prime indicates differentiation with respect to $r$.  In
general, for an assigned $\Omega$ smaller than the mass-shedding limit
the two velocities $V^{bound}$ and $V^{free}$ are different, and in
particular
\[
V^{bound}  < V^{free}.
\]
When $\Omega$ increases, the two velocities converge to the same
limit; when this limit is reached the fluid element on the surface
will not be bound anymore and the star will start loosing matter from
the equator. The value of $\Omega$ for which $V^{bound} = V^{free}$ is
the the mass-shedding limit and will be indicated as $\Omega_{ms}$.
It should be stressed that $V^{bound}$ and $V^{free}$ are computed at
the equatorial radius of the rotating configuration, i.e.
\be
\label{equaradiu}
R^{rot}=R+\xi_0(R)-\frac{1}{2}\xi_2(R).
\ee
In order to apply this procedure, we need to construct sequences of
stellar models with constant baryonic mass and variable angular
velocity.

%%%%%%%%%%%%%%%%%%%%%%%%%%%%%%%%%%%%%%%%%%%%%%%%%%%%%%%%%%%%%%%%%%%%%%%%%%%%%
\subsection{The algorithm to construct a sequence of constant baryonic mass solutions}
\label{subsection:barmass}
%%%%%%%%%%%%%%%%%%%%%%%%%%%%%%%%%%%%%%%%%%%%%%%%%%%%%%%%%%%%%%%%%%%%%%%%%%%%%
The baryonic mass of the star is defined as
\be
\label{mmbar}
M_{bar}=\int_{t=const}~ \sqrt{-g}~u^t~\epsilon_0~d^3x ,
\ee
where $g$ is the determinant of the 4-dimensional metric, 
$d^3x$ is the volume element on a $t=constant$ hypersurface,
and $\epsilon_0(r)=m_N n_{bar}(r)$ is the rest mass-energy density,
not to be confused with the energy density $\epsilon(r)$.  Since all
known interactions conserve the baryon number, $\epsilon_0$ is a
conserved quantity, i.e.
\be
(\epsilon_0u^\mu)_{;\mu}=0\,.
\ee
The expansion of $M_{bar}$  in powers of $\Omega$ is
\begin{equation}
M_{bar}=M_{bar}^{(0)}+\delta M_{bar}+O(\Omega^4)
\end{equation}
where 
\be
\label{massbar1}
M_{bar}^{(0)}=4\pi\int_0^R~dr\left(1-\frac{2M(r)}{r}\right)^{-1/2}
r^2\epsilon_0(r)
\ee
and
\beq
&&\delta M_{bar}=4\pi\int_0^Rr^2 dr~
\left(1-\frac{2M(r)}{r}\right)^{-1/2}\\
\nonumber
&&\cdot\left\{
\left[1+\frac{m_0(r)}{r-2M(r)}+\frac{1}{3}r^2
[\Omega-\omega(r)]^2e^{-\nu(r)}\right]
\epsilon_0(r)\right.\\
\nonumber
&&\left.+\frac{d\epsilon_0}{dP}(\epsilon+P)\delta p_0(r)\right\}\,.
\eeq
In the following  we shall briefly compare 
the procedure developed  by Hartle in \cite{hartle1} to construct constant baryonic mass
sequences and the procedure we use.
%*****************************************************************************
\subsubsection{Hartle's procedure}
%*****************************************************************************
\begin{enumerate}
\item{}
Given  a central energy density $\epsilon(r=0)$ and 
an assigned EOS, the  TOV equations are solved  to find
the non rotating configuration; the model is  identified by a  baryonic mass
${M}_{bar}=\bar{M}$.
\item{} For an assigned value  of the  angular velocity $\Omega$
the equations of stellar structure are solved to order $\Omega^2$,
imposing that the correction to the pressure, $\delta p_0 (r=0)$  
is different from zero.
The value of $\delta p_0(r=0)$ is then changed until the sought 
baryonic mass is reached.
\end{enumerate}

%*****************************************************************************
\subsubsection{Our procedure}
%*****************************************************************************
Our procedure to generate sequences with constant baryonic mass is the following.
\begin{enumerate}
\item{} Same as point 1 of Hartle's procedure.
\item{} For an assigned value  of  $\Omega$
we solve the equations of stellar structure up  to order $\Omega^3$,
for the different values of $\epsilon(r=0)$, by imposing that 
$\delta p_0(r=0)= 0$. 
Among these models we select  the one with baryonic mass $\bar{M}$.
\end{enumerate}
If, in addition, we want to compute the mass-shedding velocity, having computed all metric
functions for a given $\Omega$ we evaluate $V^{bound}$ and $V^{free}$. 
If $V^{bound}  = V^{free}$ we have reached  the mass-shedding limit and we stop the procedure.
If $V^{bound}  < V^{free}$ we choose a higher value of $\Omega$ and go to point  2.
This procedure can be applied to rapidly rotating stars and
to models of stars close to the maximum mass.

Hartle's procedure presents a problem  when applied to rapidly rotating stars;
indeed if the rotation rate is sufficiently high
the change in the central energy density $\delta\epsilon(r=0)$ can be so large
that treating this change as a perturbation is inappropriate and produces a large error in
all quantities. With our approach, instead, this change is treated as
a background quantity (we vary $\epsilon(r=0)$), and the results are much 
more accurate (as can be seen by the comparison with the ``exact"
integrations of \cite{cook2},\cite{EN} discussed in section \ref{section:sec3}).
A second problem arises when the mass of the star is close to
the maximum mass (where we mean the maximum mass along a constant $\Omega$ sequence).
Since $M_{bar}=M_{bar}(\Omega,\epsilon_{c})$, where 
\[
\epsilon_{c}\equiv \epsilon(r=0) +\delta\epsilon(r=0)
\]
it follows that
\[
\frac{\partial \epsilon_{c}}{\partial\Omega} {\Big\vert}_{M_{bar}} 
= - {\frac{\partial M_{bar}}{\partial
\Omega}{\Big\vert}_{\epsilon_{c}}}{\Big/}
\frac{\partial M_{bar}}{\partial \epsilon_{c}}{\Big\vert}_{\Omega}.
\] 
Since near the maximum mass
\[
\frac{\partial
M_{bar}}{\partial \epsilon_{c}}{\Big\vert}_{\Omega}\longrightarrow 0 ,
\]
it follows that 
\[
\frac{\partial \epsilon_{c}}{\partial\Omega}{\Big\vert}_{M_{bar}}\longrightarrow
\infty.
\]
Consequently, since in Hartle's procedure $\epsilon(r=0)$ is kept constant, 
and $\delta\epsilon(r=0)$  is treated as a perturbation, the procedure fails. 

%%%%%%%%%%%%%%%%%%%%%%%%%%%%%%%%%%%%%%%%%%%%%%%%%%%%%%%%%%%%%%%%%%%%%%%%%%%%%
\subsection{The moment of inertia of a rotating star}
\label{subsection:inerzia}
%%%%%%%%%%%%%%%%%%%%%%%%%%%%%%%%%%%%%%%%%%%%%%%%%%%%%%%%%%%%%%%%%%%%%%%%%%%%%
The angular momentum of an axisymmetric  matter distribution rotating about the
symmetry axis can be  expressed in terms of a conserved four vector, the
 angular momentum density
\[
J^\mu = T^\mu_{~\nu}~ \xi^\nu_{(\phi)},
\]
where $\xi^\nu_{(\phi)}$ is the Killing vector associated to axial
symmetry.  The conserved total angular momentum $J^{tot}$ can be found
by integrating $J^\mu$ over any spacelike hypersurface. Given the
coordinate system we use, the natural choice is a $t=const$
hypersurface, and since $\xi^\nu_{(\phi)}=\delta^\nu_{(\phi)}$
\be
\label{angmom}
J^{tot}= \int_{t=const}~\sqrt{-g}~ J^0~d^3x = 
 \int_{t=const}~ \sqrt{-g}T^t_{~\phi}~d^3x
\ee
where, as in (\ref{mmbar}), $g$ is the determinant of the 4-dimensional metric,
and $d^3x$ is the volume element on the chosen hypersurface.

Consequently, the moment of inertia of the matter distribution is 
\be
\label{defI1}
I=\frac{J}{\Omega} = 
\frac{1}{\Omega} \int_{t=const}~ \sqrt{-g}T^t_{~\phi}~d^3x\,.
\ee
Using Einstein's equations,
in (\ref{angmom})   and (\ref{defI1}) we can replace $T^t_{~\phi}$
with $\frac{1}{8 \pi} R^t_{~\phi}$ and expand in powers  $\Omega^n$.
By this procedure  it is possible to show that only terms with $n$ odd
contribute to this expansion, and that the first terms are
\be
\label{expmom}
J^{tot} = J + \delta J + O(\Omega^5)
\ee
where $J$ is the  first order contribution
\be
J= \frac{1}{6}\left[r^4 ~j(r)~\frac{d\varpi}{dr}\right]_{|r=R}
\label{mom1}
\ee
and $\delta J$ is the third order correction
\beq
&&\delta J=\frac{1}{6}\left\{
4r^3\frac{dj}{dr}
\varpi(\xi_0-\frac{1}{5}\xi_2) 
+r^4~j\frac{dw_1}{dr}
\right.\nn\\
&&\left.
+r^4~j\frac{d\varpi}{dr}
\left[
h_0+\frac{m_0}{r-2M}+
\frac{4(v_{2}-h_{2})}{5}
\right.\right.\nn\\
&&\left.\left.
-\frac{1}{5}\left(h_2+\frac{m_2}{r-2M}\right)\right]
\right\}_{|r=R} .
\label{mom2}
\eeq
In eqs. (\ref{mom1}) and   (\ref{mom2})  the functions $\varpi$ and $j(r)$
are defined as follows
\beq
\label{varphij}
&& \varpi(r)= \Omega - \omega(r),\\
&&j(r) = e^{-\nu(r)/2}\sqrt{1-\frac{2 M(r)}{r}}.
\nonumber
\eeq
Thus, the corrections to the moment of inertia are
\beq
&&I=I^{(0)} + \delta I + O(\Omega^4),\\
&&I^{(0)}= \frac{J}{\Omega}\,,\qquad\quad
\delta I=\frac{\delta J}{\Omega} .
\label{inertia2}
\eeq

It should be noted that  the  angular momentum defined in (\ref{angmom})
coincides with that defined using the asymptotic behaviour of the metric 
as in \cite{MTW} (Ch. 19).
In appendix A we shall show how to find $\delta J$ from the asymptotic behaviour of the metric
function $w_1$.

%%%%%%%%%%%%%%%%%%%%%%%%%%%%%%%%%%%%%%%%%%%%%%%%%%%%%%%%%%%%%%%%%%%%%%%%%%%%%
\section{A comparison with non-perturbative results}\label{section:sec3}
%%%%%%%%%%%%%%%%%%%%%%%%%%%%%%%%%%%%%%%%%%%%%%%%%%%%%%%%%%%%%%%%%%%%%%%%%%%%%
In order to verify to what extent the perturbative scheme correctly
describes the physical properties of a rapidly rotating neutron star,
we compare the results of the integration of the equations of
stellar structure expanded to order $\Omega^3$ with those of ref.
\cite{cook2} and \cite{EN}, where the fully non linear Einstein equations 
have been integrated. 

To perform the comparison we 
study stellar sequences with constant baryonic mass and
variable spin frequency $\nu=\Omega/2\pi,$
and we choose three EOS labelled as follows:\\
AU - which results from an
approach based on nuclear many body theory \cite{WFF}\\
APR2 - (APRb in \cite{EN}) which takes into account recent
results on nuclear many body theory \cite{AP,APR}\\
L - obtained from relativistic mean field theory \cite{PS}.\\
AU and L have been used both in 
\cite{cook2} and in \cite{EN}; the results of the two papers
agree to  better than 1\% for both EOS, except that for the
estimate of the moment of inertia for the EOS L, 
for which the relative error  is, at most, of the order of 1.3 \%.
APR2 has been used in \cite{EN} only.
Since we interpolate the EOS tables with the same linear
interpolation as in \cite{cook2} (in \cite{EN} a 5th-order polynomial has been used),
in the following tables
our results for AU and L will be compared  with those of \cite{cook2}.

For each EOS we consider two stellar sequences with constant baryonic mass and
varying angular velocity:\\
Sequence A, such that the gravitational mass of the non rotating 
configuration is $M=1.4\,M_\odot$.\\
Sequence B,  that corresponds to the  maximum mass.\\
The corresponding values of the baryonic mass are given in table \ref{tablea}.

%TTTTTTTTTTTTTTTTTTTTTTTTTTTTTTTTTTTTTTTTTTTTTTTTTTTTTTTTTTTTTTTTTT
\begin{table}[htbp]
\caption{The baryonic masses of the stellar models used to compare
the results of the perturbative approach with those found by integrating
the exact Einstein equations are given for the three considered EOS. The data
in column 2 and 3 correspond  to a non rotating star 
with gravitational mass $M=1.4\,M_\odot$  and $M_{max}$, respectively.}
\begin{center}
\begin{tabular}{|c|c|c|}
\hline
& \multicolumn{2}{|c|}{$M_{bar}/M_\odot $}\\
\hline
AU & 1.58 & 2.64\\
\hline
APR2 & 1.55& 2.69\\
\hline
L&1.52& 3.16\\
\hline
\end{tabular}
\end{center}
\label{tablea}
\end{table}
%TTTTTTTTTTTTTTTTTTTTTTTTTTTTTTTTTTTTTTTTTTTTTTTTTTTTTTTTTTTTTTTTTT
The values of $\nu$ are chosen as in  \cite{cook2} and \cite{EN}.
%TTTTTTTTTTTTTTTTTTTTTTTTTTTTTTTTTTTTTTTTTTTTTTTTTTTTTTTTTTTTTTTTTT
\begin{table}[htbp]
\caption{Stellar parameters for the sequence A (see text).
Data are computed  along a sequence
of stellar models with  constant baryonic mass corresponding to a non-rotating
configuration of mass $M=1.4\,M_\odot$ and 
varying spin frequency $\nu=\Omega/2\pi$
(given in KHz in column 2).
$\epsilon_c$ is the central density in units of
$\epsilon_*=10^{15}g/cm^3$; the moment of inertia $I$ is in units of
$I_*=10^{45}g\,cm^2$, the equatorial radius $R_{eq}$ is in km.  For
each value of $\nu$ and for each quantity, we give the relative error
$\Delta$.  $M_{grav}$ is the gravitational mass in solar mass units.
In the last column the ratio of the spin frequency and the
mass-shedding frequency (as computed in \cite{cook2} and \cite{EN}) is given.}
%TTTTTTTTTTTTTTTTTTTTTTTTTTTTTTTTTTTTTTTTTTTTTTTTTTTTTTTTTTTTTTTTTT
\begin{center}
\begin{tabular}{|c|c|c|c|c|c|c|}
\hline
\multicolumn{7}{|c|}{EOS AU ~($M_{bar}=1.58\,M_\odot$)}\\
\hline
& $\nu$ (KHz) & $\epsilon_c/\epsilon_*$ & $I/I_*$ & $M_{grav}/M_\odot$ &
$R_{eq}$ (Km) & $\nu/\nu_{ms}$ \\
\hline
CST & 0 & 1.206 & 1.160 & 1.400 & 10.40 &0\\
BFGM & 0 & 1.204 & 1.154 & 1.395 & 10.39 &\\
$\Delta$ & - & 0.2\% & 0.5\% & 0.4\% & 0.1\%& \\
\hline 
CST & 0.476 & 1.192 & 1.191 & 1.403 & 10.59 &0.379\\
BFGM & 0.476 & 1.190 & 1.182 & 1.398 & 10.56 &\\
$\Delta$ & - & 0.2\% & 0.8\% & 0.4\% & 0.3\%& \\
\hline
CST & 0.896 & 1.148 & 1.292 & 1.412 & 11.22 &0.713\\
BFGM & 0.896 & 1.155 & 1.254 & 1.405 & 11.02 &\\
$\Delta$ & - & 0.6\% & 2.9\% & 0.5\% & 1.8\%& \\
\hline
CST & 1.082 & 1.111 & 1.386 & 1.412 & 11.87 &0.861\\
BFGM & 1.082 & 1.132 & 1.298 & 1.410 & 11.34 &\\
$\Delta$ & - & 1.9\% & 6.3\% & 0.1\% & 4.5\%& \\
\hline
CST & 1.257 & 1.048 & 1.566 & 1.432 & 14.44 &1.000\\
BFGM & 1.257 & 1.108 & 1.344 & 1.414 & 11.73 &\\
$\Delta$ & - & 5.7\% & 14.2\% & 1.3\% & 18.8\% &\\
\hline
\hline
\multicolumn{7}{|c|}{EOS APR2 ~($M_{bar}=1.55\,M_\odot$)}\\
\hline
%& $\nu$ (KHz) & $\epsilon_c/\epsilon_*$ & $I/I_*$ & $M_{grav}/M_\odot$ &
%$R_{eq}$ (Km) & $\nu/\nu_{ms}$ \\
%\hline
BS & 0 & 0.995 & - & 1.403 & 11.55 &0\\
BFGM & 0 & 0.988 & 1.310 & 1.389 & 11.58 &\\
$\Delta$ & - & 0.7\% & - & 1.0\% & 0.3\%& \\
\hline
BS &  0.554  & 0.970 & 1.396 & 1.409 & 11.99 &0.532\\
BFGM &  0.554 & 0.964 & 1.373 & 1.394 & 11.97&\\
$\Delta$ & - & 0.6\% & 1.6\% & 1.0\% & 0.2\%& \\
\hline
BS & 0.644  & 0.960 & 1.425 & 1.411 & 12.18 &0.619\\
BFGM & 0.644 & 0.956 &  1.395 & 1.396 & 12.12 &\\
$\Delta$ & - & 0.4\% & 2.1\% & 1.0\% & 0.5\%& \\
\hline
BS & 0.879 & 0.920 & 1.551 &1.418 &13.05&0.844\\
BFGM & 0.879& 0.929& 1.466 & 1.401 & 12.63 &\\
$\Delta$ & - & 1.0\% & 5.5\% & 1.2\% & 3.2\% &\\
\hline
BS &  1.041 & 0.870 & 1.737 & 1.428 & 15.04&1.000\\
BFGM & 1.041& 0.905 & 1.526 & 1.405 & 13.13&\\
$\Delta$ & - & 4.0\% & 12.1\% & 1.6\% & 13.0\% &\\
\hline
\hline
\multicolumn{7}{|c|}{EOS L ~($M_{bar}=1.52\,M_\odot$)}\\
\hline
%& $\nu$ (KHz) & $\epsilon_c/\epsilon_*$ & $I/I_*$ & $M_{grav}/M_\odot$ &
%$R_{eq}$ (Km) & $\nu/\nu_{ms}$ \\
%\hline
CST & 0 & 0.433 & 2.127 & 1.400 & 14.99 &0\\
BFGM  & 0 & 0.440 & 2.120 & 1.402 & 14.99 &\\
$\Delta$ & - & 1.6\% & 0.3\% & 0.1\% & 0.0\% &\\
\hline
CST & 0.283 & 0.427 & 2.194 & 1.402 & 15.30 &0.395\\
BFGM  & 0.283 & 0.433 & 2.182 & 1.404 & 15.27 &\\
$\Delta$ & - & 1.4\% & 0.5\% & 0.1\% & 0.2\% &\\
\hline
CST & 0.505 & 0.411 & 2.378 & 1.407 & 16.22 &0.705\\
BFGM  & 0.505 & 0.419 & 2.307 & 1.408 & 15.95 &\\
$\Delta$ & - & 1.9\% & 3.0\% & 0.1\% & 1.7\% &\\
\hline
CST & 0.609 & 0.398 & 2.548 & 1.412 & 17.17 &0.851\\
BFGM  & 0.609 & 0.410 & 2.385 & 1.410 & 16.43 &\\
$\Delta$ & - & 3.0\% & 6.4\% & 0.1\% & 4.3\% &\\
\hline
CST  & 0.716 & 0.377 & 2.881 & 1.419 & 21.25 &1.000\\
BFGM  & 0.716 & 0.400 & 2.473 & 1.414 & 17.06 &\\
$\Delta$ & - & 6.1\% & 14.2\% & 0.4\% & 19.7\% &\\
\hline
\end{tabular}
\end{center}
\label{table1_4}
\end{table}
%TTTTTTTTTTTTTTTTTTTTTTTTTTTTTTTTTTTTTTTTTTTTTTTTTTTTTTTTTTTTTTTTTT

%TTTTTTTTTTTTTTTTTTTTTTTTTTTTTTTTTTTTTTTTTTTTTTTTTTTTTTTTTTTTTTTTTT
\begin{table}[htbp]
\caption{The stellar parameters for the sequence B, corresponding to the 
maximum mass (see text) are given as in table \ref{table1_4}.}
\begin{center}
\begin{tabular}{|c|c|c|c|c|c|c|}
\hline
\multicolumn{7}{|c|}{EOS AU ~($M_{bar}=2.64\,M_\odot$)}\\
\hline
& $\nu$ (KHz) & $\epsilon_c/\epsilon_*$ & $I/I_*$ & $M_{grav}/M_\odot$ &
$R_{eq}$ (Km) & $\nu/\nu_{ms}$ \\
\hline
CST & 0 & 3.020 & 1.982 & 2.133 & 9.41 &0\\
NOI & 0 & 3.020 & 1.964 & 2.126 & 9.39 &\\
$\Delta$ & -&0.0\% & 0.9\% & 0.3\% & 0.2\% & \\
\hline 
CST & 0.602 & 2.547 & 2.077 &2.141 & 9.74 &0.357\\
NOI & 0.602 & 2.566 & 2.058 &2.135 & 9.71 &\\
$\Delta$ & - & 0.7\% & 0.9\% & 0.3\% & 0.3\%& \\
\hline
CST & 1.174 & 2.101 & 2.281 & 2.170 & 10.40 &0.697\\
NOI & 1.174 & 2.159 & 2.218 & 2.158  & 10.19&\\
$\Delta$ & - &2.8\% &2.8 \% &0.6 \% &2.0 \%& \\
\hline
CST & 1.481 & 1.813 & 2.525 & 2.201 & 11.20 &0.879\\
NOI & 1.481 & 1.968 & 2.344 & 2.178 &  10.58 &\\
$\Delta$ & - & 8.5\% & 7.2\% & 1.0\% &6.0\%& \\
\hline
CST & 1.625  & 1.637 &2.766 & 2.227 & 12.10 &0.964\\
NOI & 1.625 & 1.881 & 2.416 & 2.187 & 10.82 &\\
$\Delta$ & - & 15.0\% & 12.7\% &1.8\% & 10.6\% &\\
\hline
CST & 1.684 & 1.532 & 2.974 &2.246 & 13.66 &1.000\\
NOI & 1.684 & 1.844 & 2.448 &2.192 & 10.93 &\\
$\Delta$ & - & 20.0\%&17.7\%& 2.4\%& 20.0 \% &\\
\hline
\hline
\multicolumn{7}{|c|}{EOS APR2 ~($M_{bar}=2.69\,M_\odot$)}\\
\hline
%& $\nu$ (KHz) & $\epsilon_c/\epsilon_*$ & $I/I_*$ & $M_{grav}/M_\odot$ &
%$R_{eq}$ (Km) & $\nu/\nu_{ms}$ \\
%\hline
BS & 0 &  2.600 & - & 2.205 &10.12&0\\
BFGM & 0 & 2.748 & 2.219  & 2.202  & 10.03&\\
$\Delta$ & - & 5.7\% & - & 0.1\% & 0.9\%& \\
\hline 
BS & 0.609  & 2.300 & 2.352 & 2.217 & 10.45 &0.408\\
BFGM & 0.609& 2.292 & 2.347 & 2.212 & 10.44 &\\
$\Delta$ & - & 0.3\% & 0.2\% & 0.2\% & 0.1\%& \\
\hline
BS & 1.081 & 1.900 &2.593 & 2.243 & 11.21 &0.723\\
BFGM & 1.081 & 1.958 &2.532 &  2.234 & 10.99&\\
$\Delta$ & - & 3.0\% & 2.4\% & 0.4\% & 2.0\% &\\
\hline
BS &1.188 & 1.800 & 2.687 & 2.253 & 11.50 &0.795\\
BFGM &1.188 & 1.884&2.589& 2.241& 11.16 &\\
$\Delta$ & - & 4.6\% & 3.6\% & 0.5\% & 3.0\% &\\
\hline
BS & 1.284 &1.700 & 2.799 &2.263 &11.86 &0.859\\
BFGM &1.284 & 1.819& 2.645& 2.248& 11.33&\\
$\Delta$ & - & 12.0\% & 5.5\% & 0.7\% & 4.5\% &\\
\hline
BS &  1.494 & 1.400 & 3.337 & 2.307 & 14.17 &1.000\\
BFGM & 1.494 &1.679 & 2.788 & 2.264 & 11.79 &\\
$\Delta$ & - & 20.0\% & 18.8\% & 2.0\% & 17.0\% &\\
\hline
\hline
\multicolumn{7}{|c|}{EOS L ~($M_{bar}=3.16\,M_\odot$)}\\
\hline
%& $\nu$ (KHz) & $\epsilon_c/\epsilon_*$ & $I/I_*$ & $M_{grav}/M_\odot$ &
%$R_{eq}$ (Km) & $\nu/\nu_{ms}$ \\
%\hline
CST & 0 & 1.470 & 4.676 & 2.700 & 13.70 &0\\
BFGM & 0 &1.486 & 4.530&  2.662 & 13.63 &\\
$\Delta$ & - & 1.6\% & 3.0\% & 1.4\% & 0.5\%& \\
\hline 
CST & 0.352&1.201&4.959&2.706&14.20&0.341\\
BFGM & 0.352  &1.220  &4.786& 2.668 &14.09 &\\
$\Delta$ & - & 1.6\% & 3.5\% & 1.4\% & 0.8\%& \\
\hline
CST & 0.714&0.955 &5.554&2.733 &15.24&0.692\\
BFGM &0.714&0.990 &5.211 &2.689 &14.84&\\
$\Delta$ & - & 3.6\% &6.0 \% & 1.6\% & 2.6\%& \\
\hline
CST &0.909 &0.802&6.292&2.765& 16.57&0.880\\
BFGM &0.909&  0.896 &5.545&2.707&15.46&\\
$\Delta$ & - & 8.4\% & 11.8\% &2.1\% &6.7\%&\\
\hline
CST &0.997&0.710&7.024& 2.792& 18.05&0.966 \\
BFGM &0.997&0.852&5.724&2.715&15.82&\\
$\Delta$ & - & 20.0\% & 18.5\% & 2.8\% & 12.4\% &\\
\hline
CST &1.032&0.655&7.659&2.813&20.66&1.000\\
BFGM &1.032&0.835&5.800&2.720&15.98&\\
$\Delta$ & - & 27.0\% &24\% & 3.3\% & 23.0\%&\\
\hline
\end{tabular}
\end{center}
\label{table_M_max}
\end{table}
%TTTTTTTTTTTTTTTTTTTTTTTTTTTTTTTTTTTTTTTTTTTTTTTTTTTTTTTTTTTTTTTTTT

The results of the comparison are summarized in Tables \ref{table1_4}
(sequence A) and \ref{table_M_max} (sequence B), where we show the stellar
parameters for different values of the spin frequency. The lines
labelled CST and BS  refer, respectively, to the data found
by integrating the exact equations of stellar structure
in \cite{cook2} and in  \cite{EN}; BFGM refer to our results  found
by integrating the equations
of stellar structure perturbed to order $\Omega^3$ for the same
models and EOS. 
The spin frequency $\nu$ is given in column 2,
the central density $\epsilon_c$  in column 3,
the moment of inertia $I$ in column 4, the gravitational mass $M_{grav}$ in column 5,
the equatorial radius $R_{eq}$ in column 6,
the ratio between the spin frequency and the
mass-shedding frequency (as computed in \cite{cook2} and \cite{EN}) in column 7.
For each value of $\nu$ and for each quantity, we give the relative error
$\Delta$.  

We stress that, as explained in Section II,
the quantities $\epsilon_c,$ $M_{grav}$ and $R_{eq}$  
are expanded in powers $\Omega^n$ with $n$ even, 
therefore, as far as our calculations are concerned, they are computed up to order $\Omega^2$.
Conversely, the angular momentum  $J^{tot}$ is expanded in odd powers of 
$\Omega$, and we compute it to order $\Omega^3$. 
The moment of inertia $I$ (see eq. \ref{defI1}) is the angular momentum divided by 
$\Omega$, therefore in this paper it is the sum of two contributions: 
$I^{(0)}$, of order $\Omega^0$ 
coming  from $J$ computed at order $\Omega$, and 
$\delta I$, of order $\Omega^2$
coming  from $\delta J$ computed at order $\Omega^3$; in this sense we refer to $I$ as to a
quantity computed up to 3rd-order in $\Omega$.

It should be noted that a difference of the order of a percent between
quantities computed by integrating the exact or the perturbed
equations can be attributed essentially to interpolation of the EOS
that are given in a tabulated form. The discrepancies are indeed of that
order when $\nu=0$, with the only exception of $\epsilon_c$
of the maximum mass star for  APR2. The larger
discrepancy found in this case ($\Delta = 5.7 \%$) can be explained
noting that the APR2 model yields a $M(\epsilon_c)$ curve featuring
an extended plateau around $M=M_{max}$. For example, a $7 \%$ change
of the central density, from $2.65 \times 10^{15} g/cm^3$ to
$2.85 \times 10^{15} g/cm^3$, only leads to a variation of the fifth
digit in the value of the mass.

From Table \ref{table1_4} 
we see that for the stars of sequence A
a critical quantity in the comparison bewteen the exact
and perturbative approach is the equatorial radius of the star;
indeed, while the relative error $\Delta =
(R_{exact}-R_{perturb.})/R_{exact}$ is smaller than $4.5 \%$ for
$\nu \lappreq 0.85~\nu_{ms}$, it becomes of the order
of $19-20 \%$ at mass-shedding. The reason of this discrepancy is that
when the spin frequency approaches the mass-shedding limit the
numerical integration of the exact equations shows that a cusp forms
on the equatorial plane, from which the star starts loosing matter
\cite{bonazzola}. This cusp is hardly reproduced by a perturbative
expansion, and in any event it cannot be reproduced by truncating the
expansion to order $\Omega^3$.  The physical quantity which is more
sensitive to the error on the radius at mass-shedding is the moment of
inertia ($\sim 14\%$). From Table \ref{table1_4}
it also emerges that the agreement between exact calculations and 3rd-order 
perturbative approach is better for the EOS APR2, which is stiffer than 
AU and L.

For the maximum mass sequences B, Table \ref{table_M_max} shows that
the discrepancies are larger; 
the largest error is on the central density, and the reason is the following. In order 
to construct a constant baryonic mass sequence, for a fixed value of 
$\nu$ the  central density is changed until the required mass is reached; since
approaching the maximum mass
\be
\lim_{M\rightarrow M_{max}}\frac{\partial\epsilon_c}{\partial M_{bar}}=\infty,
\ee
when $M=M_{max}$ the determination of $\epsilon_c$ becomes less accurate.

In Table \ref{massshedding} we show the mass-shedding frequency
computed by the perturbative and the exact approach 
for the sequences A and B: the values we find are systematically higher than those found
with the exact codes.
The discrepancy has to be attributed to the inaccuracy of the perturbative 
approach near mass-shedding as explained above.
%TTTTTTTTTTTTTTTTTTTTTTTTTTTTTTTTTTTTTTTTTTTTTTTTTTTTTTTTTTTTTTTTTT
\begin{table}[htbp]
\caption{The mass-shedding spin frequencies are given for sequences A and B (see text)
and for the considered EOS.}
\begin{center}
\begin{tabular}{|c|c|c|c|}
\hline
\multicolumn{4}{|c|}{$\nu_{ms}$ in KHz}\\
\hline
\hline
\multicolumn{4}{|c|}{Sequence A}\\
\hline
 & EOS AU & EOS L & EOS APR2\\ 
\hline
BFGM  & 1.533 & 0.876 & 1.299\\
\hline
CST & 1.257 & 0.716 & - \\ 
\hline
BS  & 1.237 & 0.728 & 1.041\\
\hline
$\Delta_{CST}$& $22.0\%$& $22.3\%$ & - \\
\hline
$\Delta_{BS}$& $23.9\%$& $20.3\%$ & $24.8\%$\\
\hline
\hline
\multicolumn{4}{|c|}{Sequence B}\\
\hline
 & EOS AU & EOS L & EOS APR2\\ 
\hline
BFGM  & 2.057 & 1.277 & 1.842\\
\hline
CST & 1.685 & 1.032 &  - \\ 
\hline
BS  & 1.682 & 1.022 & 1.494 \\
\hline
$\Delta_{CST}$& $22.1\%$& $23.7\%$ & - \\
\hline
$\Delta_{BS}$& $22.3\%$& $24.9\%$ & $23.3\%$\\
\hline
\end{tabular}
\end{center}
\label{massshedding}
\end{table}
%TTTTTTTTTTTTTTTTTTTTTTTTTTTTTTTTTTTTTTTTTTTTTTTTTTTTTTTTTTTTTTTTTT

However, it should be stressed that the rotation rate of known
pulsars is much lower than the mass-shedding limit computed for most
equation of state considered in the literature; 
for instance, the rotation rate of  the fastest isolated pulsar observed so far 
is $\nu \sim 641$ Hz, and unless we are
specifically interested in the mass-shedding problem, the results given in Tables
\ref{table1_4} and \ref{table_M_max} show
that the perturbative approach can be used to study the structural
properties of the observed rapidly rotating neutron stars.  
Indeed, for $\nu \sim  641$Hz we find
that the results of the perturbative and exact calculations for the EOS AU and APR2
differ by at most $\sim 2\%$ for all computed quantities, even for the maximum mass.

For the EOS L, differences are larger, mainly because
$\nu=641$ Hz is closer  to the  mass-shedding frequency than for the other EOS, 
and the perturbative approach becomes less accurate
for the reasons explained above. In addition,
unlike the EOS AU and APR2, the EOS L is tabulated  with very few
points, expecially at high densities (only four points for $\epsilon_c > 10^{15}~ g/cm^3$), 
and this introduces further differences which depend on the interpolation scheme.

The results of Hartle's approach have been compared to the exact results by 
several authors 
\cite{weberglen2},\cite{weberglen3},\cite{marco}.
In all these papers the perturbative approach is developed  up to order $\Omega^2$.

\cite{weberglen2} focuses on the calculation
of the mass-shedding frequency for the maximum mass configuration
$\Omega_{lim}$, which is computed for a number of EOS
to see whether the scaling law defined in \cite{hanszud} and based  on
the results of the integration of the exact equations,
is satisfied. The authors found mass-shedding frequencies  systematically larger 
by $\sim 10-15 \%$ than those obtained from the ``exact'' scaling law.

In \cite{weberglen3}, Hartle's procedure was applied to construct 
models of stars rotating at mass-shedding for three values of the mass 
and for EOS different from those we consider. 
The results were compared with those of \cite{Fried} where the exact 
equations were integrated for the same EOS, finding that
Hartle's approach leads to results
compatible to exact calculations down to rotational periods of the order
of $0.5$ ms. However, this result was subsequently disclaimed
in \cite{bonazzola1}, where it was shown that
the equatorial radius computed by the exact codes  of \cite{Fried} was 
sistematically underestimated, leading to larger values of $\Omega_{ms}$.
For this reason the agreement between \cite{weberglen3} and \cite{Fried} was good.
The conclusion of \cite{bonazzola1} is that
Hartle's method to order $\Omega^2$ cannot be considered very accurate 
in configurations approaching the mass-shedding limit, as we
also find.

Finally, in \cite{marco} 
the second order Hartle's approach is applied to construct sequencies of stellar models with
constant $M_{bar}$ and varying $\Omega$ for the same five EOS used to integrate the exact codes
of \cite{EN}. 
They used a procedure different from ours, matching  the angular momentum and 
gravitational mass computed by the two approaches varying $\nu$ and $\epsilon_c.$
Therefore in their approach $J$ (computed to order $\Omega$)
 and $M_{grav}$ are equal by construction.
The comparison between the perturbative and the exact approaches 
is focused on the quadrupole moment $Q,$ which is the coefficient of the $r^{-3}
P_2(\cos\theta)$ term in Newtonian potential \cite{hartlet}.  To order $\Omega^2,$
$Q$ is
\be
Q=\frac{J^2}{M}+\frac{8}{5}KM^3
\ee
where $K$ is defined in eq. (\ref{extgen2}).
They found a relative error which depends on the EOS and increases with
the rotation rate. For rates of the order of  the fastest pulsar they found 
errors of the order of $ 20 \%$.
Our calculations of $Q$ for the EOS APR2, AU and L that are also used in 
\cite{EN} show a much better agreement with the exact results, with
a relative  error smaller than $2\%$ for $M_{grav}=1.4~M_\odot$.

%%%%%%%%%%%%%%%%%%%%%%%%%%%%%%%%%%%%%%%%%%%%%%%%%%%%%%%%%%%%%%%%%%%%%%%%%%%%%

%%%%%%%%%%%%%%%%%%%%%%%%%%%%%%%%%%%%%%%%%%%%%%%%%%%%%%%%%%%%%%%%%%%%%%%%%%%%%
%%%%%%%%%%%%%%%%%%%%%%%%%%%%%%%%%%%%%%%%%%%%%%%%%%%%%%%%%%%%%%%%%%%%%%%%%%%%%
\section{The EOS}
\label{section:eos}
%%%%%%%%%%%%%%%%%%%%%%%%%%%%%%%%%%%%%%%%%%%%%%%%%%%%%%%%%%%%%%%%%%%%%%%%%%%%%
%%%%%%%%%%%%%%%%%%%%%%%%%%%%%%%%%%%%%%%%%%%%%%%%%%%%%%%%%%%%%%%%%%%%%%%%%%%%%

As stated in Section I, in this work we have employed different models of 
the EOS of neutron 
star matter in the region of supranuclear density, corresponding to the 
star core, while the description of the outer and inner crust provided by the EOS
of Refs. \cite{BPS} and \cite{PRL}, respectively, has been kept fixed.
The four EOS considered in the core have been obtained within two approches, nuclear many
body theory (NMBT) and relativistic mean field theory (RMFT), based on different 
dynamical assumptions and different formalisms.

The NMBT approach rests on the premise that nuclear matter can be described as 
a collection of nonrelativistic nucleons, interacting through phenomenological 
two- and three-body forces. While suffering from the obvious limitations inherent 
in its nonrelativistic nature, this approach is made very attractive by its feature
of being strongly constrained by the data. The two-nucleon potential provides a 
quantitative account of the properties of the
two-nucleon system, i.e. deuteron properties and $\sim$ 4000 accurately measured 
nucleon-nucleon scattering phase shifts \cite{WSS}, while the inclusion of the 
three-nucleon potential allows one to reproduce the binding energies of the 
three-nucleon bound states \cite{PPCPW}, as well the empirical equilibrium 
properties of symmetric nuclear matter \cite{APR}. Calculations of the energies 
of the ground and low-lying excited states of nuclei with mass number 
$A\leq10$ \cite{WP}, whose results are in excellent agreement with the experimental 
values, have also shown that NMBT has a remarkable predictive power.

Within RMFT, based on the formalism of relativistic quantum field theory, nucleons 
are described as Dirac fermions interacting through meson exchange.
In its simplest implementation the dynamics is described by a scalar and a vector 
meson field \cite{QHD1}. This approach is very elegant, but leads to a set
of equations of motion that turns out to be only tractable in the mean field
approximation, i.e. treating the meson fields as classical fields. The meson 
masses and coupling constant are then fixed fitting the equilibrium properties of
nuclear matter.

Both NMBT and RMFT can be generalized to take into account the possible appearance of 
strange baryons, produced in weak interaction processes that may become energetically 
favourable at high density. However, the inclusion of baryons other than protons and 
neutrons entails a large uncertainty, as little is known of their interactions and 
the available models are only loosely constrained by few data. 

Of the four EOS we consider, those referred to as APR2 
(also used in  Section \ref{section:sec3} to compare our results with those of
non-perturbative approaches), BSS1 and BSS2 are based
on NMBT, while the one labelled G240 has been obtained from RMFT. The APR2 and
BSS1 EOS include nucleons only, and their differences are mainly ascribable 
to the approximate treatment of the three-nucleon interactions in the BSS1 
model and to the presence of relativistic boost corrections in the APR2 model.
The BSS2 model is a generalization of BSS1 including the hyperons $\Sigma^-$ and
$\Lambda^0$. Finally, the G240 EOS includes the full baryon octet, consisting 
of protons, neutrons and $\Sigma^{0,\pm}$, $\Lambda^0$ and $\Xi^{\pm}$. 

The non rotating neutron star configurations corresponding to the APR2 and BSS1 EOS
exhibit similar mass-radius relations, with a maximum gravitational mass 
$\gsim$ 2 M$_\odot$.
The appearance of strange baryons makes the EOS softer, thus leading to sizable
changes in the mass-radius relation. This feature is clearly visible in the case
of both the BSS2 and G240 EOS \cite{omarleo}, and leads to maximum masses 
of $\sim$ 1.22 M$_\odot$ and $\sim$ 1.55 M$_\odot$, respectively.

Comparison with the experimental determinations of the masses of neutron stars located 
in binary radio pulsar systems, yielding a narrow distribution centered at 1.35 $\pm$0.04 
M$_\odot$ \cite{masses}, suggests that the dynamics underlying the BSS2 model leads to
too soft an EOS. On the other hand, all other EOS appear to be compatible with the
mass of the most massive neutron star listed in Ref. \cite{masses} 
($\sim$ 1.56 M$_\odot$).

%%%%%%%%%%%%%%%%%%%%%%%%%%%%%%%%%%%%%%%%%%%%%%%%%%%%%%%%%%%%%%%%%%%%%%%%%%%%%%%%%%%%%
%%%%%%%%%%%%%%%%%%%%%%%%%%%%%%%%%%%%%%%%%%%%%%%%%%%%%%%%%%%%%%%%%%%%%%%%%%%%%
\section{Results}
\label{section:results}
%%%%%%%%%%%%%%%%%%%%%%%%%%%%%%%%%%%%%%%%%%%%%%%%%%%%%%%%%%%%%%%%%%%%%%%%%%%%%
Using the above EOS to describe matter in the crust and in the stellar core, and
using the procedure developed in section \ref{section:sec2}, 
the equations of stellar structure perturbed  to order $\Omega^3$
have been integrated to construct sequencies of stellar 
configurations with constant baryonic mass and varying angular velocity.
In our study we  consider six values of $M_{bar},$ plus the maximum mass model for each EOS.
The gravitational mass is
\be
M_{grav}=M+\delta M_{grav}+O(\Omega^4),
\ee
where
\be
M=4\pi \int_0^R~ r^2~\epsilon(r)~dr
\ee
is the gravitational mass of the non rotating configuration
and $\delta M_{grav}$ is given by eq. (\ref{masschange});
$ M_{grav}$  depends on the EOS and on the angular
velocity. In Table \ref{mbartab} for each EOS we give the values of $M$ for the non
rotating model that corresponds to the selected $M_{bar}$
%TTTTTTTTTTTTTTTTTTTTTTTTTTTTTTTTTTTTTTTTTTTTTTTTTTTTTTTTTTTTTTTTTTTTTT
\begin{table}[htbp]
\label{mbartab}
\caption{For each value of the baryonic mass in column 1, 
the corresponding values of the gravitational mass of the non rotating configuration
are tabulated for the selected EOS.}
\begin{center}
\begin{tabular}{|c|c|c|c|c|}
\hline
\multicolumn{5}{|c|}{$M/\msun$}\\
\hline
 $M_{bar}/\msun$ & APR2& BBS1 & BBS2& G240 \\
\hline
1.31&1.19 &1.18&1.19&1.20\\
1.50&1.35 &1.34& - &1.37\\
1.56&1.39 &1.38& - &1.41\\
1.68&1.49 &1.47& - &1.51\\
2.10&1.80 &1.79& - & -\\
2.35&2.00 &1.97& - & -\\
\hline
\end{tabular}
\end{center}
\end{table}
%TTTTTTTTTTTTTTTTTTTTTTTTTTTTTTTTTTTTTTTTTTTTTTTTTTTTTTTTTTTTTTTTTTTTTT

%TTTTTTTTTTTTTTTTTTTTTTTTTTTTTTTTTTTTTTTTTTTTTTTTTTTTTTTTTTTTTTTTTTTTTTTTTT
\begin{table}[htbp]
\caption{ Maximum masses of the non rotating configurations.}
\begin{center}
\begin{tabular}{|c|c|c|}
\hline
 EOS & $M^{max}_{bar}/M_\odot$& $M^{max}/M_\odot$\\
\hline
APR2&2.70  &2.20\\
BBS1&2.41   &2.01\\
BBS2& 1.74 &1.22\\
G240& 1.35 &1.55\\
\hline
\end{tabular}
\label{maxtab}
\end{center}
\end{table}
%TTTTTTTTTTTTTTTTTTTTTTTTTTTTTTTTTTTTTTTTTTTTTTTTTTTTTTTTTTTTTTTTTTTTTT

The maximum masses (baryonic and gravitational) of the non rotating configurations 
are given in Table \ref{maxtab} for the selected EOS.

\begin{figure}[htbp]
\begin{center}
\leavevmode
\centerline{\epsfig{file=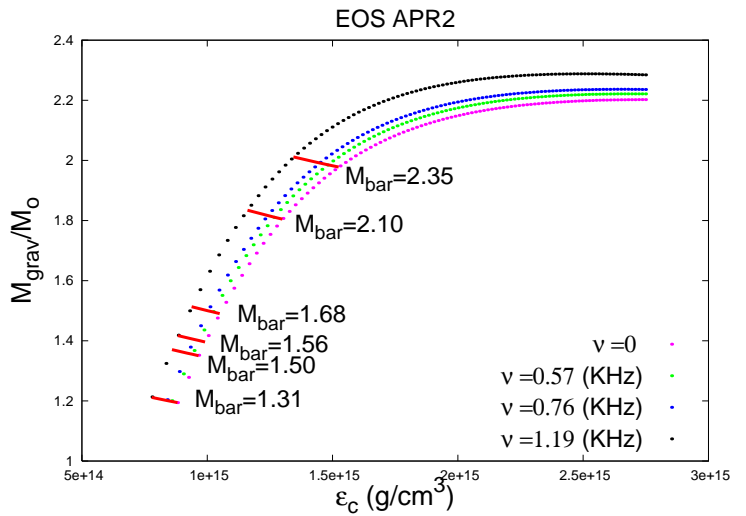,height=9.6cm,angle=270}}
\caption{(Color online) The gravitational mass for the EOS APR2 is plotted
as a function of the central density for different values of
the spin frequency, evidentiating the curves with $M_{bar}=const$.}
\label{FIG1}
\end{center}
\end{figure}

An example of how the gravitational mass changes due to rotation is shown in Fig. \ref{FIG1},
where we plot $M_{grav}$ (in solar mass units) as a function of the central 
density for different values of the spin frequency for the EOS APR2, evidentiating 
the sequencies with $M_{bar}=const$.

%%%%%%%%%%%%%%%%%%%%%%%%%%%%%%%%%%%%%%%%%%%%%%%%%%%%%%%%%%%%%%%%%%%%%%%%%
%%%%%%%%%%%%%%%%%%%%%%%%%%%%%%%%%%%%%%%%%%%%%%%%%%%%%%%%%%%%%%%%%%%%%%%%%

%ssssssssssssssssssssssssssssssssssssssssssssssssssssssssssssssssssssss
\subsection{Mass-shedding limit}
\begin{figure}[htbp]
\begin{center}
\leavevmode
\centerline{\epsfig{file=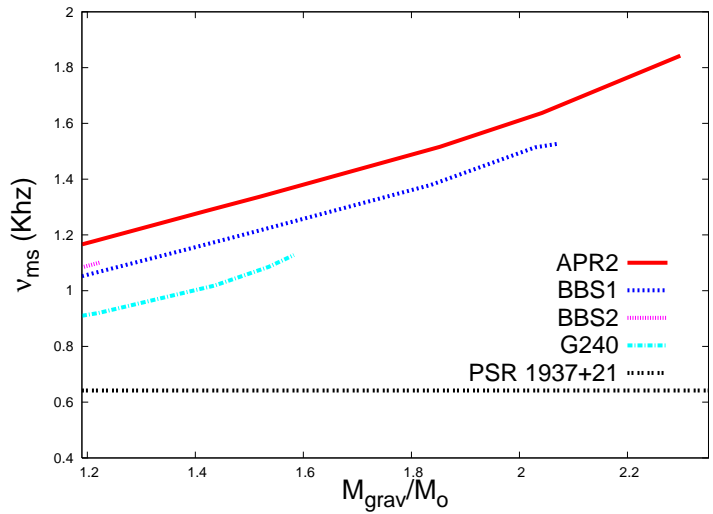,height=9.6cm,angle=270}}
\caption{(Color online) The  mass-shedding frequency $\nu_{ms}$
is plotted for each EOS  as a function of the gravitational mass. The horizontal line corresponds
to the spin frequency of PSR 1937+21, which is the fastest isolated pulsar observed 
so far and whose mass is
presently unknown.}
\label{FIG2}
\end{center}
\end{figure}

%ssssssssssssssssssssssssssssssssssssssssssssssssssssssssssssssssssssss
In Fig. \ref{FIG2} we plot the  mass-shedding frequency, $\nu_{ms},$
as a function of the gravitational mass for each EOS.
The dashed horizontal line is the  frequency of  PSR 1937+21, the fastest isolated
pulsar observed so far,  the mass of which is presently unknown.
%%%%%%%%%%%%%%%%%%%%%%%%%%%%%%%%%%%%%%%%%%%%%%%%%%%%%%%%%%%%%%%%%%%%%%%%%
%%%%%%%%%%%%%%%%%%%%%%%%%%%%%%%%%%%%%%%%%%%%%%%%%%%%%%%%%%%%%%%%%%%%%%%%%

We see that the  values of $\nu_{ms}$ for G240 are sistematically
lower than those of the other EOS, while those  for APR2  are sistematically higher.
This behaviour can be understood from the data given in
Table \ref{tab6} where, for each EOS,  we tabulate the
ratio of the gravitational mass of the  star rotating at mass-shedding 
to the equatorial radius, $M^{ms}_{grav}/R^{ms}_{eq},$ for three values of the baryonic mass.
For a given mass, the ordering of $\nu_{ms}$ in Fig. \ref{FIG2} corresponds to the same
ordering of $M^{ms}_{grav}/R^{ms}_{eq}$ in Table \ref{tab6}, showing that
more compact stars admit a higher mass-shedding limit.
%%%%%%%%%%%%%%%%%%%%%%%%%%%%%%%%%%%%%%%%%%%%%%%%%%%%%%%%%%%%%%%%%%%%%%%%%
\begin{table}[htbp]
\caption{ For each EOS  and for three values of the baryonic mass,
we tabulate the ratio of the gravitational mass of the  star rotating at mass-shedding 
to the equatorial radius, $M^{ms}_{grav}/R^{ms}_{eq}.$ 
In the last three columns, we give  the ratio $M/R$ for the 
non rotating star with the same baryonic mass.}
\begin{center}
\begin{tabular}{|c|c|c|c|c|c|c|}
\hline 
&\multicolumn{3}{|c|}{$\nu=\nu_{ms}$}
&\multicolumn{3}{|c|}{$\nu=0$}\\
\hline
&\multicolumn{3}{|c|}{$M^{ms}_{grav}/R^{ms}_{eq}$}
&\multicolumn{3}{|c|}{$M/R$}\\
\hline
$M_{bar}/M_{\odot}$&
1.31& 1.56&2.35&
1.31& 1.56&2.35\\
\hline
G240 &  0.106 & 0.127 &  - & 0.131 &  0.163  & -  \\ 
\hline
BBS1 &0.114 &  0.136 &  0.210  & 0.143 &  0.169 & 0.267 \\
\hline
BBS2 & 0.117 & - & - & 0.154 & - & - \\
\hline 
APR2 & 0.124 & 0.147 &  0.223&  0.152 & 0.179 &  0.268  \\
\hline
\end{tabular}
\end{center}
\label{tab6}
\end{table}
%%%%%%%%%%%%%%%%%%%%%%%%%%%%%%%%%%%%%%%%%%%%%%%%%%%%%%%%%%%%%%%%%%%%%%%%%

In the last three column of Table \ref{tab6} we also give the 
ratio $M/R$ of the non rotating star with the same baryonic mass, and it
is interesting to see that rotation affects the compactness of a star 
in a way which depends on the EOS.
For instance, for $M_{bar}= 1.31 \msun$ (for which all EOS admit a stable configuration), 
the non rotating star with EOS BBS2 ($M/R=0.154$) is more compact 
than that with EOS APR2 ($M/R=0.152$),
while near mass-shedding the compactness of the BBS2  star ($M^{ms}_{grav}/R^{ms}_{eq}=0.117$) 
is smaller than that of the APR2 star ($M^{ms}_{grav}/R^{ms}_{eq}=0.124$).

%%%%%%%%%%%%%%%%%%%%%%%%%%%%%%%%%%%%%%%%%%%%%%%%%%%%%%%%%%%%%%%%%%%%%%%%%
%\begin{figure*}
\begin{figure}[htbp]
\begin{center}
\leavevmode
%\subfigure[EOS APR2]{\input{file=FIG3a.ps}}
%\subfigure[EOS BBS2]{\input{file=FIG3b.ps}}
\centerline{\epsfig{file=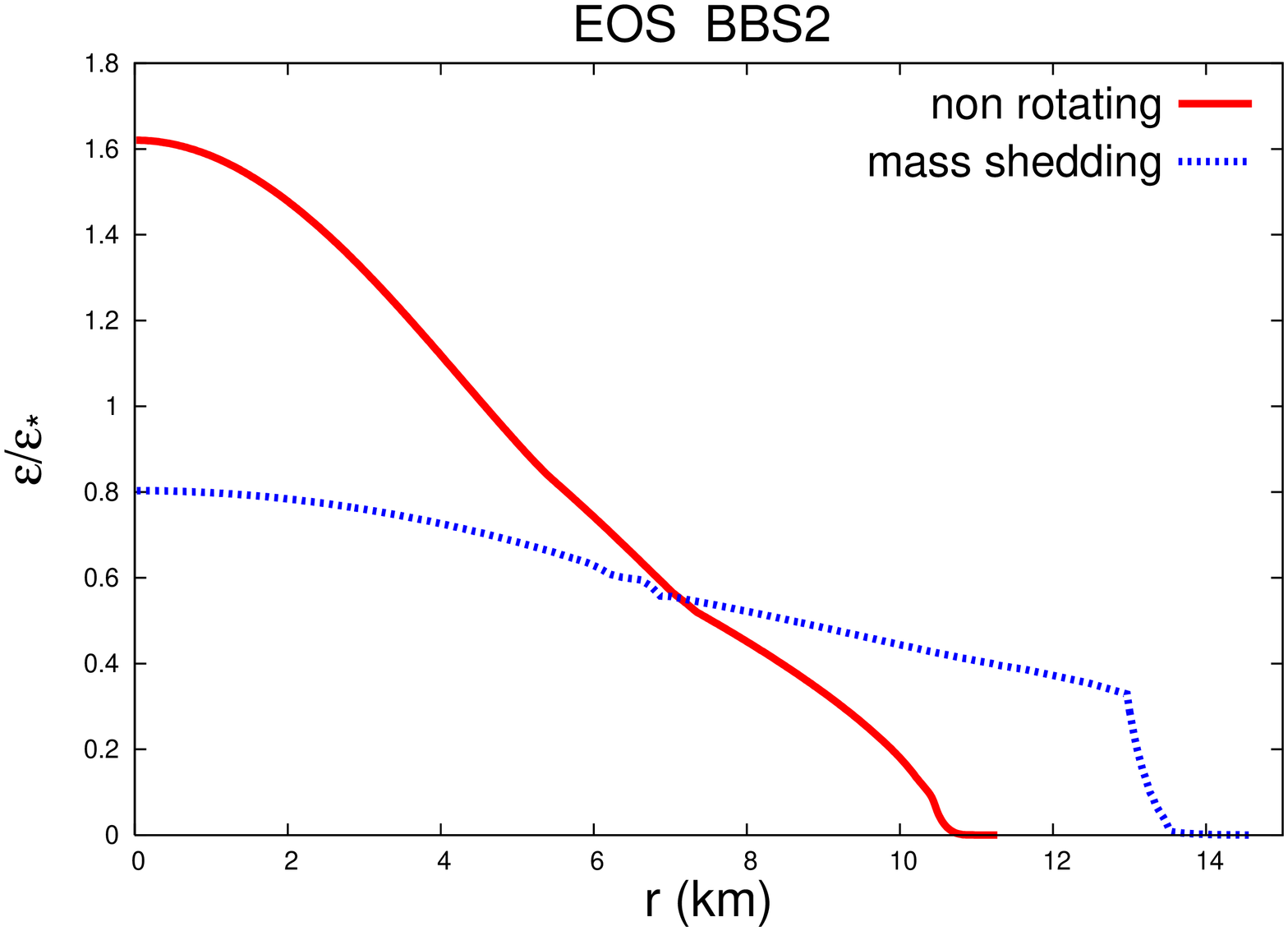,height=9.6cm,angle=270}}
\vskip 12pt
\centerline{\epsfig{file=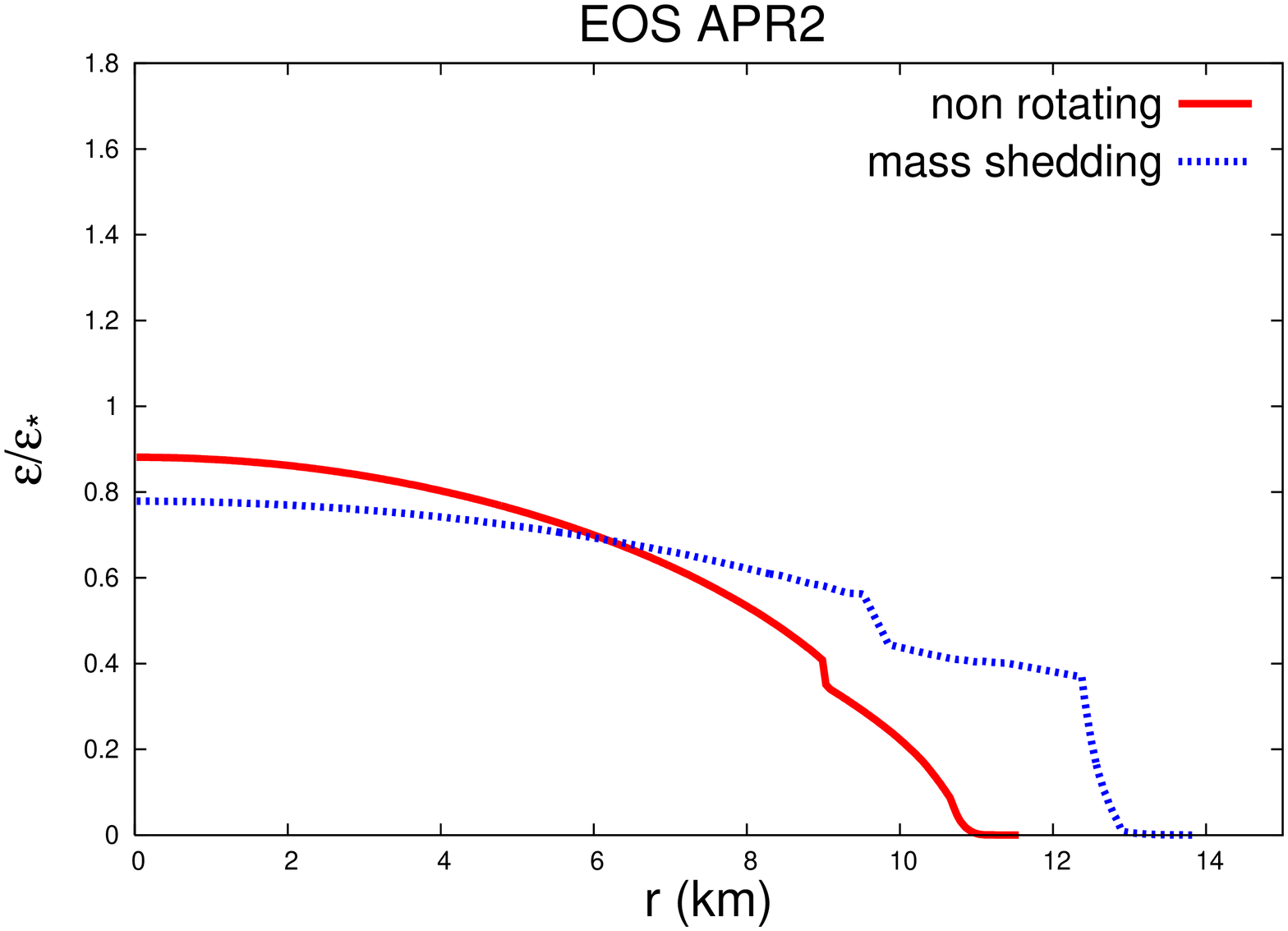,height=9.6cm,angle=270}}
\caption{
(Color online) The energy density distribution inside the star is plotted 
as a function of the radial distance for the  EOS APR2 and BBS2, and for $M_{bar}=1.31$.
$\epsilon$ is given in units of $\epsilon_*=10^{15}~g/cm^3$.
In both cases the continuous line refers to the non rotating configuration, the
dotted line to the star rotating at mass-shedding.
}
\label{FIG3}
\end{center}
\end{figure}
%\end{figure*}
%%%%%%%%%%%%%%%%%%%%%%%%%%%%%%%%%%%%%%%%%%%%%%%%%%%%%%%%%%%%%%%%%%%%%%%%%
To further clarify this behaviour in Fig. \ref{FIG3} we plot
the energy density inside the star as a function of the radial distance for these two EOS
in the non rotating case and at mass-shedding. It is useful to remind that the main
difference between the EOS APR2 and BBS2 is that in the core of the BBS2 star there are
hyperons. 
%%%%%%%%%%%%%%%%%%%%%%%%%%%%%%%%%%%%%%%%%%%%%%%%%%%%%%%%%%%%%%%%%%%%%%%%%
\begin{figure*}[htbp]
\begin{center}
\leavevmode
\centerline{\epsfig{file=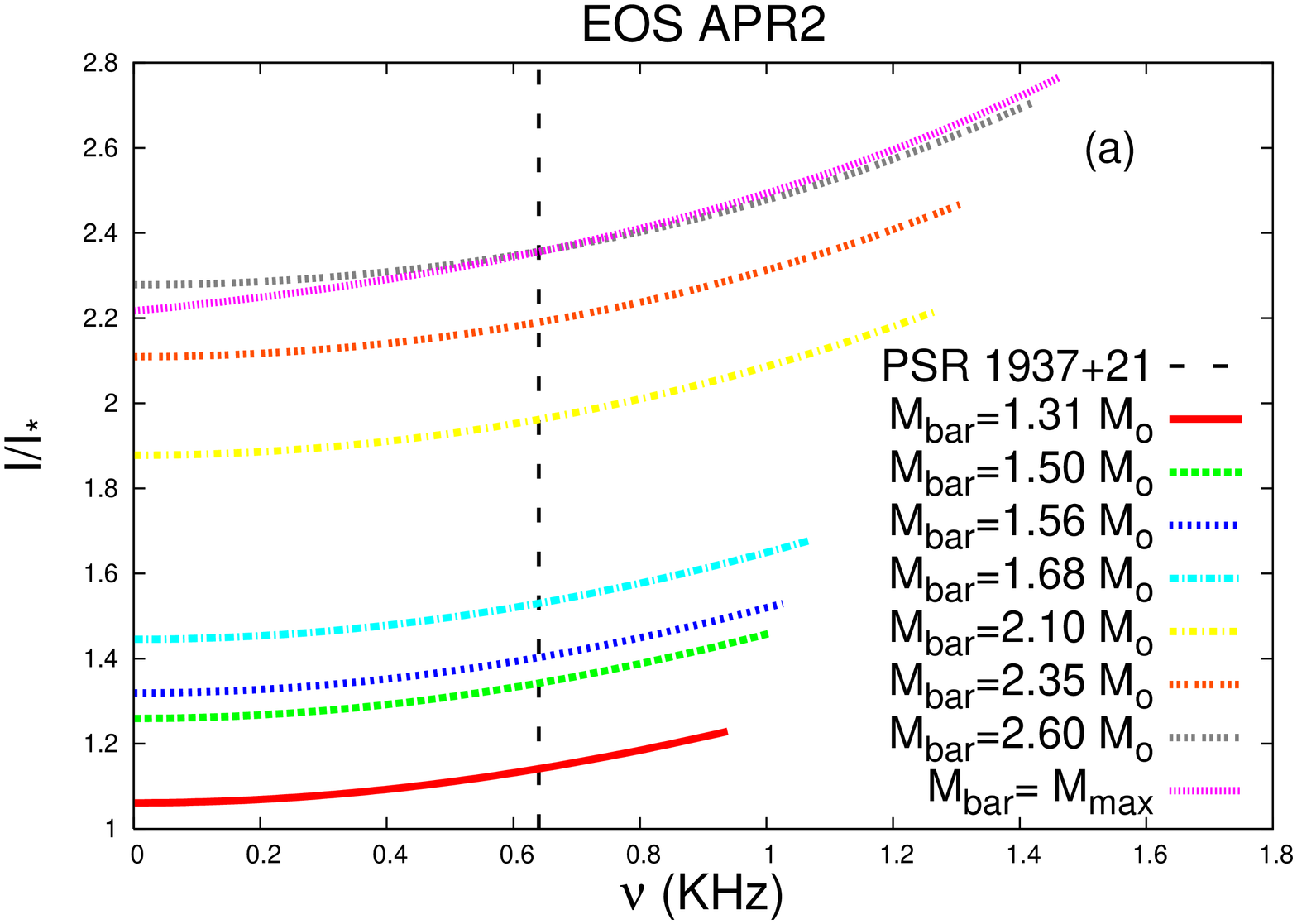,height=9.6cm,angle=270},\epsfig{file=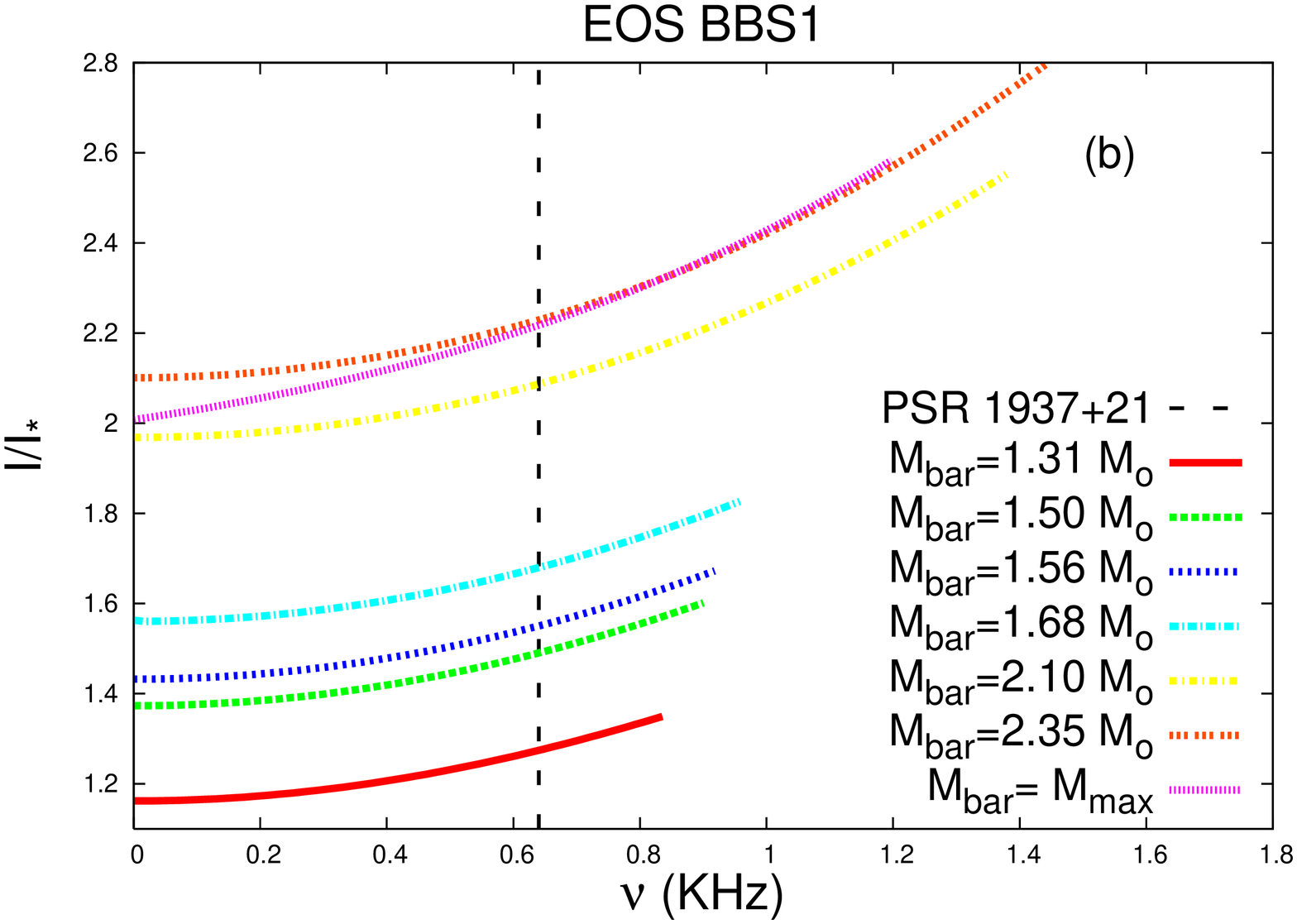,height=9.6cm,angle=270}}
 \vskip 12pt
\centerline{\epsfig{file=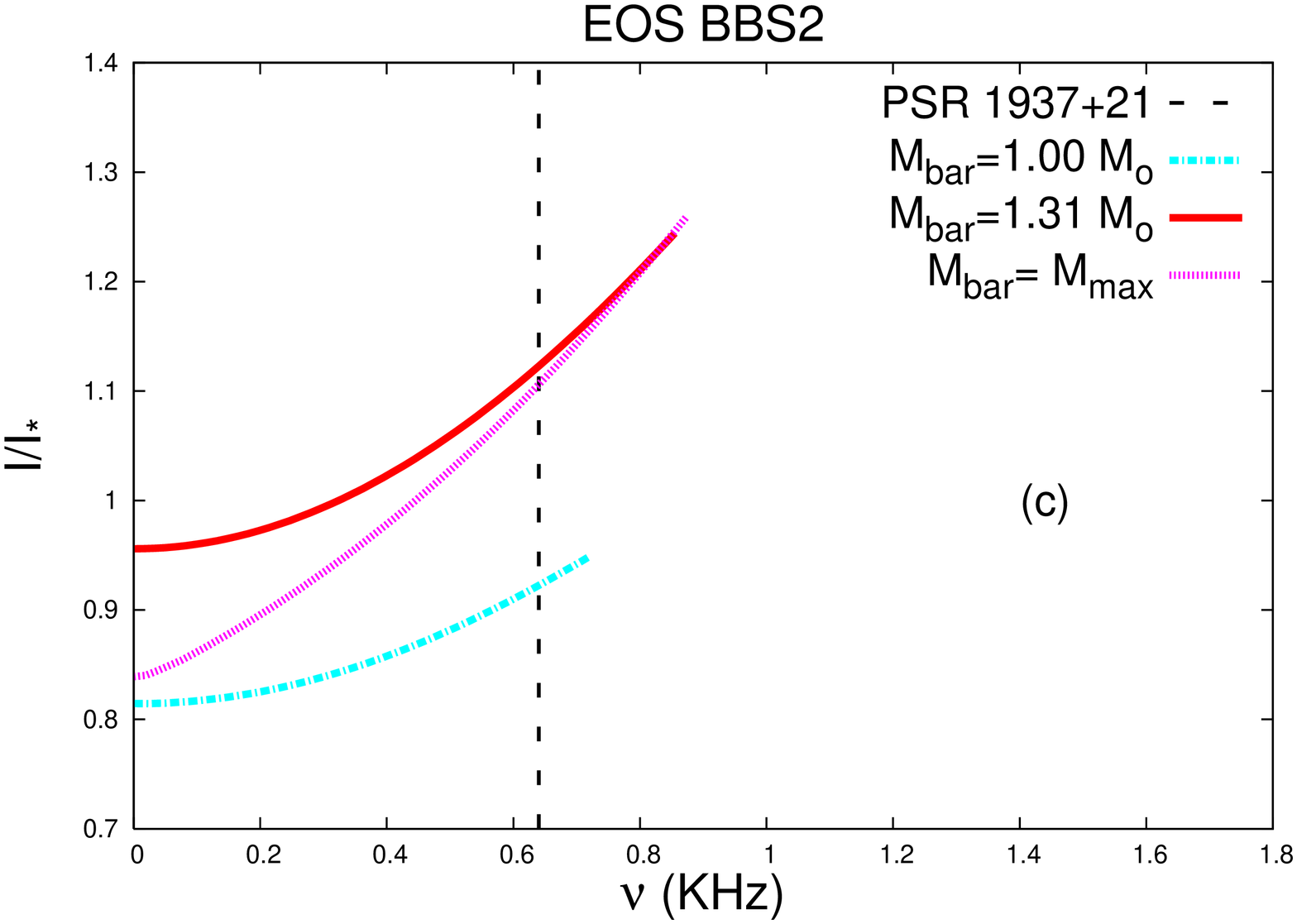,height=9.6cm,angle=270},\epsfig{file=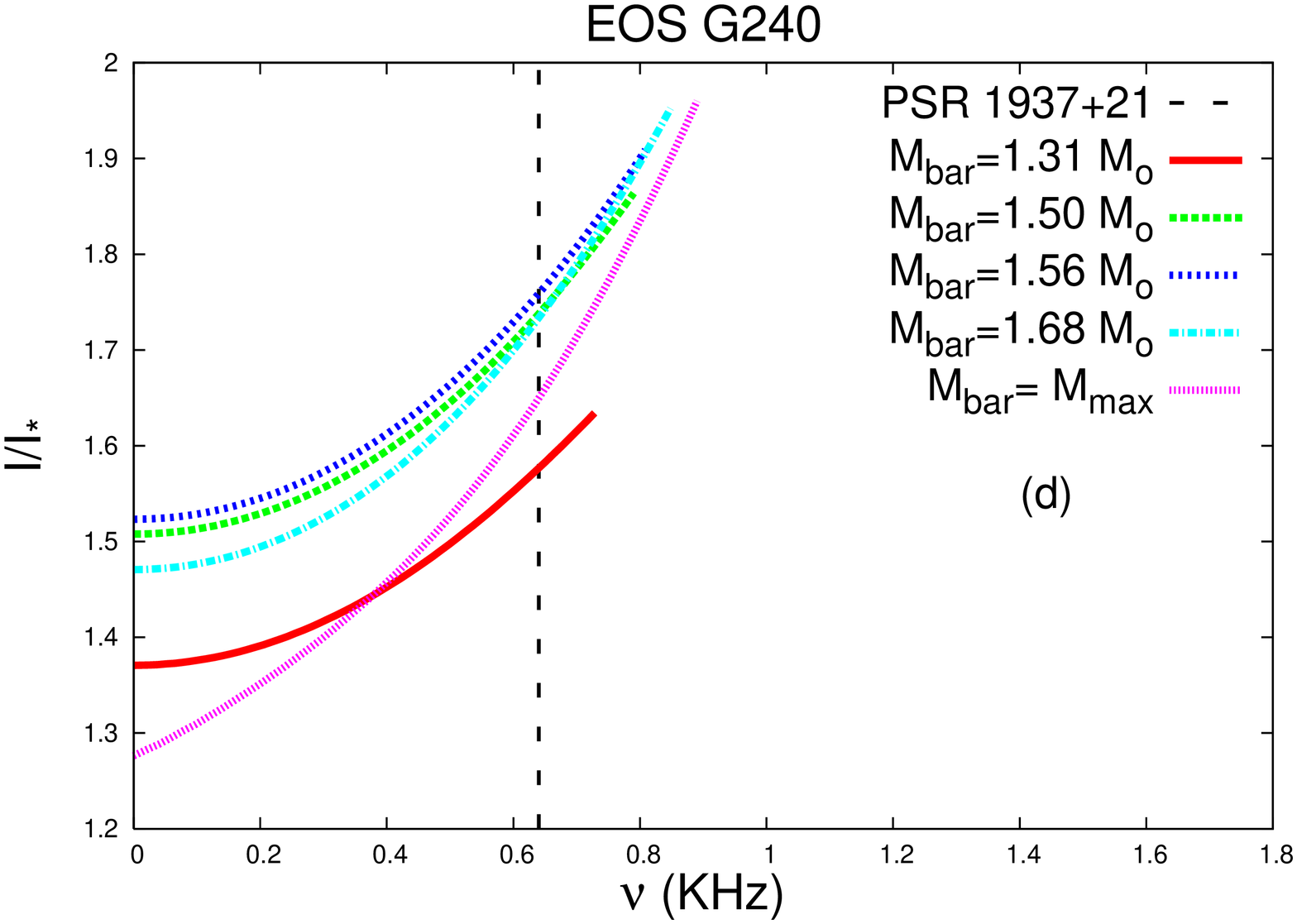,height=9.6cm,angle=270}}
 \caption{
(Color online) The moment of inertia (in units of $I_*=10^{45}g\,cm^2$) is plotted 
as a function of the spin frequency,
for the EOS APR2 (panel a), BBS1 (panel b), BBS2 (panel c) and
G240 (panel d). For each EOS we consider a few values of the baryonic mass up to the
maximum mass, and plot the data only for  $\nu \leq 80 \% ~\nu_{ms}$ (see text).
The vertical dashed line is the spin  frequency of  PSR 1937+21.
}
\label{FIG4}
\end{center}
%\end{figure}
\end{figure*}
%%%%%%%%%%%%%%%%%%%%%%%%%%%%%%%%%%%%%%%%%%%%%%%%%%%%%%%%%%%%%%%%%%%%%%%%%
From the upper panel of Fig. \ref{FIG3} we see that when the BBS2 star
does not rotate hyperons are highly concentrated in the core, making
the central density very large; however, since the EOS is soft, when
the star rotates fastly matter is allowed to distribute throughout
the star.  Conversely, due to the stiffness of the APR2 EOS, in the
corresponding star (Fig. \ref{FIG3} lower panel) the matter
distribution does not seem to be so affected by rotation.
A similar analysis can be found in \cite{Fried} were different EOS were considered.

By comparing our results to order $\Omega^3$ 
with those to order $\Omega^2$  we find that the third order
contributions to the mass-shedding limit 
are actually negligible, smaller than $1\%$, independently of the EOS.

%ssssssssssssssssssssssssssssssssssssssssssssssssssssssssssssssssssssss
\subsection{Estimates of the moment of inertia}
\label{subsection:inerziaest}
%ssssssssssssssssssssssssssssssssssssssssssssssssssssssssssssssssssssss
In Fig. \ref{FIG4} we plot the moment of inertia as a function of the
spin frequency, for each EOS and for given values of the
baryonic mass up to the maximum mass.  Since we know, as discussed in
section III, that the mass-shedding frequencies estimated by our
perturbative approach are about $20 \%$ larger than those found by
integrating the exact equations, we plot the data only
for $\nu \leq 80 \% ~\nu_{ms}$, where $\nu_{ms}$ is the mass-shedding
frequency we find.  From Fig. \ref{FIG4} we see that the moment of
inertia has a non-trivial behaviour as the baryonic mass increases. 
Indeed, along a $\Omega=constant$ sequence of
stellar models, if $M_{bar}\ll M_{bar}^{max}$ then $I$
is an {\em increasing} function of $M_{bar}$, while 
as $M_{bar}$ approaches the maximum mass 
$I$ becomes a {\em decreasing} function of $M_{bar}$.  

This behaviour can be understood as follows:
$I \sim MR^2$, and while in general keeping $\Omega$ constant
as the mass increases the radius decreases, approaching the maximum mass
the mass remains nearly constant while the radius is continuosly decreasing. 
This is true for any EOS, but of course it is more evident
for softer EOS like BBS2 and G240. 

In a recent paper \cite{baum} it has been suggested that a measurement
of the moment of inertia from pulsar timing data will impose
significant constraints on the nuclear EOS.  The authors consider the
newly discovered binary double pulsar PSR J037-3039, and construct
models of the fastest pulsar ($M=1.337~\msun, ~\nu= 276.8 ~Hz$) with
different  EOS. They compute the moment of inertia assuming the star is
rigidly rotating using the fully non linear, numerical code
developed in \cite{cook2}, and compare the results with those obtained
using Hartle's perturbative approach developed at first order in the
angular velocity, i.e. they compute the quantity given in
eq. (\ref{inertia2}) which we call $I^{(0)}$. They show that the
agreement is very good, and this had to be expected since the spin
frequency of the considered star is quite low.
%%%%%%%%%%%%%%%%%%%%%%%%%%%%%%%%%%%%%%%%%%%%%%%%%%%%%%%%%%%%%%%%%%%%%%%%%
\begin{figure}[htbp]
\begin{center}
\leavevmode
\centerline{\epsfig{file=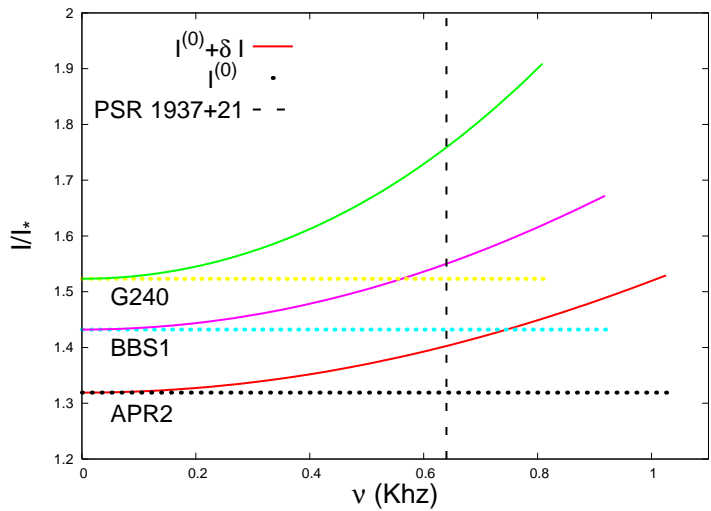,height=9.6cm,angle=270}}
\caption{(Color online) The  moment of inertia, measured in units of $I_*=10^{45}g\,cm^2$,
is plotted versus the spin frequency (in KHz) for
stars with baryonic mass $M_{bar}=1.56 \msun$. 
The continuous lines refer to $I$ computed by solving the stellar structure equations
to third order in the angular velocity, whereas the horizontal dashed lines refer to the
first order term $I^{(0)}$.
The vertical dashed line is the spin  frequency of  PSR 1937+21.
For each EOS, the data are plotted only for  $\nu \leq 80 \% ~\nu_{ms}$ (see text).
}
\label{FIG5}
\end{center}
\end{figure}
%%%%%%%%%%%%%%%%%%%%%%%%%%%%%%%%%%%%%%%%%%%%%%%%%%%%%%%%%%%%%%%%%%%%%%%%%
What we want to show now is that if we consider stars that rotate
faster, as for example PSR 1937+21, the first order approach is
inaccurate, and we need to consider the third order corrections.  To
this purpose, in Fig. \ref{FIG5} we plot for each EOS and for
$M_{bar}=1.56~\msun$ the first order term $I^{(0)}$ (dashed line) and
$I=I^{(0)} + \delta I$ (solid line) where $\delta I$ is the correction
found by solving the perturbed equations to order $\Omega^3$.  For
each EOS, the data are plotted for $\nu \leq 80 \% ~\nu_{ms}.$ The
vertical dashed line refers to the spin frequency of PSR 1937+21.  It
is clear from the figure that the third order correction becomes
relevant at frequencies higher than $\sim 0.5 ~KHz$. For instance if
$\nu=641~Hz$ the relative error $[\delta I/(I^{(0)}+\delta
I)]$ that one makes neglecting the third order corrections are $5\%$ 
for APR2, $7 \%$ for BBS1 and $13 \%$ for
G240.

It is interesting to compare $I^{(0)}$, $I=I^{(0)} + \delta I$, 
where $\delta I$ is computed using corrections of order $\Omega^3$, 
and the value of $I$ one gets by integrating the
exact equations of stellar structure. This is possible for the EOS
APR2 and the results are shown in Table \ref{table8} for 
$M_{grav}\sim 1.4\,M_\odot$ and  $\nu = 644\,Hz$, and for the maximum mass and 
$\nu = 609\,Hz$. 

The table shows that including third order is
important to adequately estimate the moment of inertia of the fastest
pulsars.
%%%%%%%%%%%%%%%%%%%%%%%%%%%%%%%%%%%%%%%%%%%%%%%%%%%%%%%%%%%%%%%%%%%%%%%%%
\begin{table}[htbp]
\caption{We compare the moment of inertia computed for the EOS
APR2 by integrating the exact equations of stellar structure (column 2) \cite{EN},
the perturbed equations to  order  $\Omega$ (column 3), and to order
$\Omega^3$ (column 5). The relative error bewteen the exact and the approximated results
are given in columns 4 and 6. 
$I$ is in units of $I_*=10^{45}g\,cm^2$.
}
\begin{center}
\begin{tabular}{|c|c|c|c|c|c|}
\hline
&$I_{exact}/I_*$& $I^{(0)}/I_*$ & $\Delta $ & $\left( I^{(0)} + \delta I\right)/I_*$ & $\Delta $\\
\hline
$M_{grav} \sim 1.4~M_\odot$&
1.425 & 1.238 & 13$\%$ & 1.395 &  2.1$\%$ \\
$\nu=644$ Hz&&&&&\\
\hline
&$I_{exact}/I_*$& $I^{(0)}/I_*$ & $\Delta $ & $\left( I^{(0)} + \delta I\right)/I_*$ & $\Delta $\\
\hline
$M_{grav}=M_{max}$&
2.352 & 2.281 & 3$\%$ & 2.347 &  0.2$\%$ \\
$\nu=609$ Hz&&&&&\\
\hline
\end{tabular}
\end{center}
\label{table8}
\end{table}
%%%%%%%%%%%%%%%%%%%%%%%%%%%%%%%%%%%%%%%%%%%%%%%%%%%%%%%%%%%%%%%%%%%%%%%%%

%%%%%%%%%%%%%%%%%%%%%%%%%%%%%%%%%%%%%%%%%%%%%%%%%%%%%%%%%%%%%%%%%%%%%%%%%
%%%%%%%%%%%%%%%%%%%%%%%%%%%%%%%%%%%%%%%%%%%%%%%%%%%%%%%%%%%%%%%%%%%%%%%%%
\section{Concluding Remarks}
%%%%%%%%%%%%%%%%%%%%%%%%%%%%%%%%%%%%%%%%%%%%%%%%%%%%%%%%%%%%%%%%%%%%%%%%%
%%%%%%%%%%%%%%%%%%%%%%%%%%%%%%%%%%%%%%%%%%%%%%%%%%%%%%%%%%%%%%%%%%%%%%%%%

In this paper we have solved the equations of stellar structure
developed to third order in the angular velocity, 
using recent EOS which model hadronic interactions 
in different ways to describe the matter in the inner core.
The stellar parameters we find have been compared, when available in the literature, 
with those found by solving the exact equations of stellar  structure.
The main conclusions we can draw from our study are the following.

It  is known  that near mass-shedding the perturbative approach 
fails to correctly reproduce the stellar properties and the reasons
are well understood.
We confirm this behaviour and give quantitative results in this limit.

However, for lower values of the  angular velocity the situation is different.
Taking as a reference the rotation rate of the fastest isolated 
pulsar observed so far, PSR 1937+21, for which  $\nu \sim 641$ Hz,
we see that at these rates the perturbative approach
allows to describe all stellar properties to an accuracy better than 
$\sim 2\%$  even for the maximum mass models, unless the EOS in the core
is very soft,  as in the case of the EOS L;
indeed, for this EOS  the mass-shedding velocity
is low and  close to the chosen rotation rate, so that the perturbative approach 
is inaccurate.

The two quantities that are affected by third order corrections are the mass-shedding velocity
and the moment of inertia. For the first, we find that the third order
corrections are actually negligible, smaller than $1\%$, independently of the EOS.

Conversely, the moment of inertia is affected by terms of order $\Omega^3$ in a significant
way; for instance, for a star with $M=1.4~M_\odot$ rotating at $\nu \sim 641$ Hz 
the third order correction is of the order of 
$\delta I/(I^{(0)}+\delta I)\sim 5\%$ for the stiffer EOS we use (APR2), and as
high as $\sim 13\%$ for the softest  (G240).

Thus, third order corrections have to be included
to study the moment of inertia  of rapidly rotating neutron stars, whereas
they are irrelevant to  estimate the mass-shedding limit.
%\newpage
%%%%%%%%%%%%%%%%%%%%%%%%%%%%%%%%%%%%%%%%%%%%%%%%%%%%%%%%%%%%%%%%%%%%%%%%%
\appendix
\section{} \label{appendixA}
%%%%%%%%%%%%%%%%%%%%%%%%%%%%%%%%%%%%%%%%%%%%%%%%%%%%%%%%%%%%%%%%%%%%%%%%%

In  this appendix  we write the equations that govern the radial part of the metric 
and  thermodynamical functions that describe the structure of a rotating  star 
to third order in the angular velocity  $\Omega$. For each function we specify the
appropriate boundary conditions and we show how to construct the solution order by
order. We give explicitely the analytic solution when available,
following and completing the work of Hartle and collaborators 
\cite{hartle,hartle1,hartlet}.
In what follows we shall refer to the functions appearing in section \ref{section:sec2}.

We recall that the structure of a non
rotating, spherical star is described by the TOV equations
\beq
\label{tovequation1}
&&\frac{dM(r)}{dr}=4\pi r^2\epsilon(r), 
\\\nn
&&\frac{d\nu(r)}{dr}=2\frac{M(r)+4\pi r^3P(r)}{r(r-2M(r))}, 
\\\nn
\label{tovequation3}
&&\frac{dP(r)}{dr}=-\frac{(P(r)+\epsilon(r))}{2}\nu(r)_{,r}.
\eeq
$\epsilon(r)$ and $P(r)$ are the energy density and pressure distribution in the non rotating
star.
Since all functions we shall consider  depend on $r$, we shall  omit this dependency throughout.\\
Hereafter, $R$ and $M(R)$ will indicate, respectively, the radius and the mass
 of the non rotating star, found by solving eqs. (\ref{tovequation1}) for an assigned EOS.

%SSSSSSSSSSSSSSSSSSSSSSSSSSSSSSSSSSSSSSSSSSSSSSSSSSSSSSSSSSSSSSSSSSSSSSSSSSS
\subsection{The first order equations}
%SSSSSSSSSSSSSSSSSSSSSSSSSSSSSSSSSSSSSSSSSSSSSSSSSSSSSSSSSSSSSSSSSSSSSSSSSSS
To  order $\Omega$ the shape of the star remains spherical and
there is only one function, $\omega$, to determine, which is responsible for the dragging of
inertial frames (cfr. the expansion in eq. \ref{oexp}).
By introducing the function $ \varpi=\Omega-\omega,$
it is easy to show that it  satisfies the following equation \cite{hartle}
\be
\frac{1}{r^4}\frac{d}{dr}(r^4j\frac{d\varpi}{dr})+\frac{4}{r}
\frac{dj}{dr}\bar{\varpi}=0,
\label{eqpai}
\ee
where $\label{defj} j(r)= e^{-\nu/2}\sqrt{1-\frac{2 M}{r}}. $
Equation (\ref{eqpai}) can be reduced to two first order equations by defining
two auxiliary functions of $r$
\be
\label{defuchi}
\chi=j\varpi,\qquad\qquad
u=r^4j\frac{d\varpi}{dr}
\ee
that satisfy
\beq
r \le R\quad&&\frac{d\chi}{dr}=\frac{u}{r^4}-\frac{4\pi ,
\nonumber
r^2(\epsilon+P)\chi}{r-2M}\\
&&\frac{du}{dr}=\frac{16\pi r^5(\epsilon+P)\chi}{r-2M} .
\label{eqpai1}
\eeq
They have to be integrated from $r=0$ to $R$ with the following boundary conditions
\be
\chi(0)=j(0)\varpi_c,\qquad\quad
u(0)=0.
\label{cond1}
\ee
These conditions follow from the behaviour of $\varpi$ 
near the origin 
\be
\label{varp}
\varpi \sim \varpi_c(1 + \varpi_2 r^2 +....),
\ee
where $\varpi_c$ is a constant.
Note that from eqs. (\ref{defuchi}) and (\ref{varp}) it fopllows that $u$ goes to
zero faster than $r^4$.

For $r \ge R,$  $P$ and $\epsilon$ vanish,  $M \equiv M(R)$, and
the solution of eqs. (\ref{eqpai1}) is
\be
\label{uchiout}
r \ge R\qquad \chi(r)=\Omega-\frac{2J}{r^3},\qquad
u(r)=6J,
\ee
where $J$ is a constant which represents the angular momentum of the star to 
first order in $\Omega$. 
The constants $\varpi_c$ and $J$ can be found by imposing that the interior and the exterior
solutions match continuously at $R,$
i.e. 
\[
\varpi_c ~\chi(R) =\Omega-\frac{2J}{R^3},\qquad\quad
u(R)= 6 J.
\]
%%%%%%%%%%%%%%%%%%%%%%%%%%%%%%%%%%%%%%%%%%%%%%%%%%%%%%%%%%%%%%%%%%%
\subsection{Second order equations}
%%%%%%%%%%%%%%%%%%%%%%%%%%%%%%%%%%%%%%%%%%%%%%%%%%%%%%%%%%%%%%%%%%%%
The second order terms affect the structure of the star in the following way:
$h_0$ and $m_0$ produce a spherical expansion,  $h_2$ and $v_2$
a quadrupole deformation.  

On the assumption that the EOS is a one parameter equation of state,
Einstein's equations admit a  first integral (cfr. \cite{hartle})
that allows to find two equations, one
relating $h_0$  and $\delta p_{0}$, the second  $h_2$ and $\delta p_{2}$.
These two equations are
\beq
\label{prearmsfe}
\bar\mu=\delta p_{0}+h_0-\frac{1}{3}e^{-\nu}r^2\varpi^2\\
\label{prearmsfe2}
0=\delta p_2+h_2+\frac{1}{3}e^{-\nu}r^2\varpi^2\,,
\eeq
where $\bar\mu$ is the correction of order $\Omega^2$ to the chemical potential
$\mu_c$
\be
\mu_c=\mu[1+\bar\mu+O(\Omega^4)]
\equiv\frac{\cal E+P}{u^t}\exp\left[-\int\frac{d\cal E}{\cal E+P}\right]
\ee
(where ${\cal E}$ and ${\cal P}$ are the energy density and pressure
in the rotating configuration)
which can be shown to be constant with respect to $r$ and $\theta$ throughout the star.

%%%%%%%%%%%%%%%%%%%%%%%%%%%%%%%%%%%%%%%%%%%%%%%%%%%%%%%%%%%%%%%%%%%
\subsubsection{Spherical expansion}
%%%%%%%%%%%%%%%%%%%%%%%%%%%%%%%%%%%%%%%%%%%%%%%%%%%%%%%%%%%%%%%%%%%%
Using eq. (\ref{prearmsfe}), the relevant equations for the spherical deformation  
can be shown to be
\beq
\nonumber
&&\frac{dm_0}{dr}=4\pi r^2\frac{d\epsilon}{dP}[\delta p_0(\epsilon+P)]+
\frac{u^2}{12r^4}+\frac{8\pi r^5(\epsilon+P)\chi^2}{3(r-2M)}
\\
\nonumber
&&\frac{d\delta p_{0}}{dr}=\frac{u^2}{12r^4(r-2M)}
\\
\nonumber
&&-\frac{m_0(1+8\pi r^2P)}{(r-2M)^2} 
-\frac{4\pi(\epsilon+P)r^2\delta p_0}{r-2M}\\
&&+\frac{2r^2\chi}{3(r-2M)}
\left[\frac{u}{r^3}+\frac{(r-3M-4\pi r^3P)\chi}{r-2M}\right]
\label{m0edeltap1}
\eeq
where $u$ and $\chi$ are defined in (\ref{defuchi}). These equations 
must be integrated inside the star from $r=0$ to $R$  with the condition
that both $m_{0}$ and $\delta p_{0}$ vanish in $r=0$. To start the numerical integration
it is useful to make use of the following expansion
\beq
\label{asint1}
&&m_{0}(r) \sim \frac{4\pi}{15}\left[\epsilon(0)+P(0) \right]
\left[ \left(\frac{d\epsilon}{dP}\right)_{r=0} + 2\right] ~\chi(0)^2~ r^5
\nonumber
\\
&&\delta p_{0}(r) \sim  \frac{1}{3} ~\chi(0)~r^2, \qquad r \rightarrow 0.
\eeq
For $r \ge R~$    $u$ and $\chi$ reduce to
(\ref{uchiout}); by integrating the first eq. (\ref{m0edeltap1}) with $P$ and  $\epsilon$
vanishing, one finds
\be
\label{m0fuori}
m_0(r)=-\frac{J^2}{r^3}+\delta M
\ee
where $\delta M$ is a constant which represents
the correction to the gravitational mass due to the spherical
expansion of the star. This quantity can be determined by imposing the continuity of the
solution at $r=R$, i.e.

\be
\delta M=m_0(R)+\frac{J^2}{R^3},
\label{deltam}
\ee
and $m_0(R)$ is found by integrating eqs. (\ref{m0edeltap1}) for $r \le R$.

Due to the sperical expansion, the pressure of each element of fluid 
changes by an amount $\delta p_0(r)$ found by integrating eqs. (\ref{m0edeltap1});
at the same time the element is radially displaced by an amount
\be
\label{rotradius}
r + \xi_0(r),
\ee 
and $\xi_0(r)$ can be found in terms of $\delta p_0(r)$  using eq. (\ref{pres0}); known $\xi_0(r)$
we can compute  the corresponding variation of the stellar radius (cfr. eq. \ref{deltaraggioA})

\be
\delta R^A=\xi_0(R)=\frac{\delta p_0(R)R(R-2M(R))}{M(R)}.
\label{deltaraggio}
\ee

%%%%%%%%%%%%%%%%%%%%%%%%%%%%%%%%%%%%%%%%%%%%%%%%%%%%%%%%%%%%%%%%%%%
\subsubsection{Quadrupole deformation}
%%%%%%%%%%%%%%%%%%%%%%%%%%%%%%%%%%%%%%%%%%%%%%%%%%%%%%%%%%%%%%%%%%%%
We shall  now solve the equations for $h_2(r)$ e $v_2(r)$, that are responsible for the
quadrupole deformation of the star:

\beq
\label{quad}
\frac{dv_2}{dr}&=&-\frac{d\nu}{dr}h_2+(\frac{1}{r}+
\frac{1}{2}\frac{d\nu}{dr})\left[
\frac{8\pi r^5(\epsilon+P)\chi^2}{3(r-2M)}+\frac{u^2}{6r^4}\right]
\\\nonumber
\frac{dh_2}{dr}&=&\left[-\frac{d\nu}{dr}+
\frac{r}{r-2M}(\frac{d\nu}{dr})^{-1}(8\pi(\epsilon+P)-
\frac{4M}{r^3})\right]h_2\\
\nonumber
&-&\frac{4v_2}{r(r-2M)}(\frac{d\nu}{dr})^{-1}+
\frac{u^2}{6r^5}\left[
\frac{1}{2}\frac{d\nu}{dr}r-\frac{1}{r-2M}(\frac{d\nu}{dr})^{-1}\right]
\\\nonumber
&+&\frac{8\pi r^5(\epsilon+P)\chi^2}{3(r-2M)}\left[
\frac{1}{2}\frac{d\nu}{dr}r+
\frac{1}{r-2M}(\frac{d\nu}{dr})^{-1}\right].
\eeq

Let us consider the solution for $r \le R$ first.

It is easy to check that a regular solution of eqs. (\ref{quad}) 
near the origin must behave as
\be
r \rightarrow 0\qquad h_2(r) \sim A r^2,\qquad v_2 \sim B r^4,
\label{asint2}
\ee
where the constants  $A$ and $B$ are related by the following expression
\be
\label{asint3}
B + 2\pi \left[\frac{1}{3}\epsilon(0)+P(0) \right] A 
= \frac{2}{3}\pi \left[\epsilon(0)+P(0) \right] \left(j(0)\varpi_c\right)^2.
\ee
Following \cite{hartle}, we write the general solution of eqs. (\ref{quad}) for 
$r \le R$ as
\beq
\label{gen1}
&&h_{2}(r)=h^P_2 + C~h^H_{2}\\\nonumber
&&v_{2}(r)=v^P_2 + C~v^H_{2}.
\eeq 
$C$ is a constant to be determined (see below),
$(h_2^P,v_2^P)$ are a particular solution found
by integrating eqs. (\ref{quad}) with the initial  conditions, say, 
$A=1$ and $B$ given  by eq.  (\ref{asint3}); 
$(h_2^H,v_2^H)$ are the solution  of the homogeneous system 
which is found by putting $\chi=0$ and $u=0$ in eqs. (\ref{quad}), numerically integrated
with the boundary condition
\beq
r \rightarrow 0
&&h^H_{2}(r)\sim r^2\\\nonumber    
&&v^H_{2}(r)\sim -2\pi~ \left[\frac{1}{3}\epsilon(0)+P(0) \right] r^4.\nonumber
\eeq

For $r \ge R$ eqs. (\ref{quad}) can be solved analytically in terms of the associated Legendre
functions of the second kind, and the solution is \cite{hartle},\cite{hartlet}
\beq
\label{extgen2}
r \ge R
&&h_2(r)=J^2(\frac{1}{M(R) r^3}+\frac{1}{r^4})+K~Q^2_2(\xi)\\\nonumber     
&&v_2(r)=-\frac{J^2}{r^4}+ K~ \frac{2M(R)}{[r(r-2M(R))]^{1/2}}Q^1_2(\xi)
\eeq
where $K$ is a constant and
\beq
\label{defq}
&&Q^2_2(\xi)=[\frac{3}{2}(\xi^2-1)log(\frac{\xi+1}{\xi-1})-
\frac{3\xi^3-5\xi}{\xi^2-1}]\\\nonumber
&&Q^2_1(\xi)=(\xi^2-1)^{1/2}[\frac{3\xi^2-2}{\xi^2-1}-
\frac{3}{2}\xi log(\frac{\xi+1}{\xi-1})].
\eeq
with
\be
\label{defxi}
\xi=\frac{r}{M(R)} -1. 
\ee
We can now determine the constants $C$ and $K$ by imposing that the solution (\ref{gen1}) and the
solution (\ref{extgen2}) are equal at $r=R;$ thus,
the functions  $h_2$ and $v_2$ are completely determined.

We can now compute the quadrupole deformation of the star. Indeed, known $h_2$ from
eq. (\ref{prearmsfe2}) we find $\delta p_2$:
\be
\label{deltap2}
\delta p_2(r) = - h_2(r)-\frac{1}{3}e^{-\nu(r)}r^2\varpi(r)^2,
\ee
and  from eq. (\ref{pres2}) we find the second correction to the displacement
\be
\label{xi2}
\xi_2(r) = - \delta p_2(r) {\Big/}  \left[\frac{1}{\epsilon+P}\frac{dP}{dr}\right].
\ee
Thus,  an element of fluid located at a given $r$ and a given $\theta$ in the 
non rotating star, when the star rotates moves to a new position identified by the same
value of $\theta$, and by a radial coordinate
\be
\label{displ1a}
\bar{r} = r + \xi_0(r)+\xi_2(r)P_2(\theta)+ O(\Omega^4),
\ee
where $\xi_0(r)$ and $\xi_2(r)$ are found as explained above.

The remaining metric function $m_2$ can be determined  by a suitable combination of Einstein's
equations, which gives
\be
\label{defm2}
m_2=(r-2M)\left[-h_{2}+\frac{8\pi r^5(\epsilon+P)\chi^2}{3(r-2M)}+
\frac{u^2}{6r^4}\right].
\ee

%%%%%%%%%%%%%%%%%%%%%%%%%%%%%%%%%%%%%%%%%%%%%%%%%%%%%%%%%%%%%%%%%%%
\subsection{Third order equations}
%%%%%%%%%%%%%%%%%%%%%%%%%%%%%%%%%%%%%%%%%%%%%%%%%%%%%%%%%%%%%%%%%%%%
The functions that describe the third order corrections are $w_1$ and $w_3,$ and
satisfy two second order differential equations with a similar structure. 
Since they do not couple, we  shall solve them independently.
%%%%%%%%%%%%%%%%%%%%%%%%%%%%%%%%%%%%%%%%%%%%%%%%%%%%%%%%%%%%%%%%%%%
\subsubsection{The equations for $w_1$}
%%%%%%%%%%%%%%%%%%%%%%%%%%%%%%%%%%%%%%%%%%%%%%%%%%%%%%%%%%%%%%%%%%%
\be
\label{eqw1}
\frac{1}{r^4}\frac{d}{dr}(r^4j\frac{dw_1(r)}{dr})+
\frac{4}{r}\frac{dj}{dr}w_1=D_0-\frac{1}{5}D_2,
\ee
where
\beq
\label{newd0}
&&r^4D_0=-u\frac{d}{dr}\left[\frac{m_{0}}{r-2M}+h_0\right]
+4r^3\chi\left[\frac{1}{j}\frac{dj}{dr}\right]
\\\nonumber
&& \cdot \left[\frac{2m_{0}}{r-2M}
+(\frac{1}{(dP/d\epsilon)}+1)\delta p_0+\frac{2r^3\chi^2}{3(r-2M)}
\right],\\
\label{newd2}
\nonumber\\\nonumber
&&\frac{r^4D_2}{5}=\frac{u}{5}\frac{d}{dr}\left[4k_2-
h_{2}-\frac{m_2}{r-2M} \right]
+(\frac{4r^3\chi}{5})\left[\frac{1}{j}\frac{dj}{dr}\right]
\\\nonumber
&&
\cdot\left[\frac{2m_2}{r-2M}
+(\frac{1}{(dP/d\epsilon)}+1)\delta p_2-\frac{2r^3\chi^2}{3(r-2M)}
\right].
\eeq
By introducing  the variables
\be
\label{defu1chi1}
\chi_1=jw_1\qquad\qquad
u_1=r^4j\frac{dw_1}{dr}
\ee
eq. (\ref{eqw1}) becomes
\beq
\label{neweqw1}
&&{\frac{d\chi_1}{dr}=\frac{u_1}{r^4}-\frac{4\pi
r^2(\epsilon+P)\chi_1}{r-2M}} ,\\\nonumber
&&{\frac{du_1}{dr}=\frac{16\pi r^5(\epsilon+P)\chi_1}{r-2M}}+r^4D_0-\frac{r^4}{5}D_2.
\eeq 
We write the general solution of this system  for $r \le R$ as
\beq
\label{gen1a}
&&\chi_1=\chi^P_1 + C~\chi^H_1\\\nonumber
&&u_1=u^P_1 + C~u_1^H.
\eeq 
$C$ is a constant,
$(\chi^P_1,u^P_1)$, are a particular solution found
by integrating eqs. (\ref{neweqw1}) with the initial  conditions 
\[
\chi^P_1 (0) =0,\qquad u^P_1(0) =0;\]
$(\chi^H_1, u_1^H)$ is the solution  of eqs. (\ref{neweqw1})
in which   $D_0$ and $D_2$ are set to zero,  numerically integrated 
with the asymptotic conditions
\beq
r \rightarrow 0
&&\chi^H_1(r)\sim 
1+\frac{8\pi}{5}\left[\epsilon(0)+P(0) \right]~r^2, \\\nonumber    
&&u_1^H(r)\sim \frac{16\pi}{5}~ \left[\epsilon(0)+P(0) \right]~r^5.
\eeq
For $r \ge R$ eqs. (\ref{neweqw1}) reduce to
\beq 
\label{extneweqw1}
&&\frac{d\chi_{1}}{dr}=\frac{u_{1}}{r^4} ,\\\nonumber
&&\frac{du_{1}}{dr}=-\frac{u}{5}\frac{d}{dr}[4k_2- \frac{6J^2}{r^4}].
\eeq
Note that, since $j(r\ge R)=1,$ for $r \ge R$ $\chi_1$ and  $w_1$ coincide (cfr. eq.
\ref{defu1chi1}).

At large distance from the source $w_1$ must behave as
\beq
\label{asinw1}
w_1(r) =\frac{2\delta J}{r^3}+0(\frac{1}{r^4}),
\eeq
where $\delta J$ is the correction to the angular momentum of the star;
this suggests to write $w_1,$ and therefore $\chi_1,$  as
\be
\label{extw1}
w_1(r) = \chi_1(r) =\frac{2\delta J}{r^3}+F(r),
\ee
where  $F(r)$ must go to zero at infinity faster than
$r^3.$ 
Consequently, when  $r \rightarrow \infty$~~ $u_1 \rightarrow - 6 \delta J$.
Thus, by direct integration of the second eq. (\ref{extneweqw1}), and using eqs. 
(\ref{uchiout}), (\ref{kexp}) and (\ref{extgen2}), we find 
\beq
\label{newsol1}
\nonumber
&&u^{ext}_{1}=-\frac{u}{5}[4k_2-\frac{6J^2}{r^4}] -6\delta J
=\frac{84J^3}{5r^4}+\frac{24J^3}{5 M(R) r^3}
\\
&&+\frac{24 J K Q^2_2}{5}-\frac{48 M(R) J K Q^1_2}{5[r(r-2M(R))]^{1/2}}
-6\delta J.
\eeq
Knowing $u^{ext}_1,$ we can now integrate the first of eqs. (\ref{extneweqw1}) which, using 
eq. (\ref{extw1}), becomes  an equation for $F(r)$, the solution
of which is
\beq
\label{effe}
&&F(r)=-\frac{12J^3}{5r^7}-\frac{4J^3}{5M r^6}+\\\nonumber
&&\frac{JK}{40M^3r^4}\left[108r^4\ln(\frac{r}{r-2M})
-288r^3M\ln(\frac{r}{r-2M})
\right.\nonumber\\
&& \left.  +33r^4- 240r^3M + 336r^2M^2+256M^3r-96M^4\right.\nonumber\\
&&\left.+192rM^3\ln(\frac{r}{r-2M})+12r^4\ln(\frac{r}{r-2M})\right]
-\frac{33JK}{40M^3},
\nonumber
\eeq
where $M\equiv M(R).$ 
This expression satisfies eq. (\ref{extneweqw1}), as it can be checked by direct
substitution; it  may be noted that it differs from the expression of $F(r)$
 given in ref. \cite{hartle} which, conversely, does not
satisfy eq.(\ref{extneweqw1}). 

In conclusion
\be
\chi_1^{ext}=\frac{2\delta J}{r^3}+F(r).
\label{chi1ext}
\ee
At this point the situation is the following.  We have integrated
eqs. (\ref{neweqw1}) for $r\le R$, and we know the solution for $u_1$
and $\chi_1$ up to the constant $C$ (eq.~ \ref{gen1a}); we have found
the solution for $r\ge R$ for $u_1$ (eq.\ref{newsol1}) and for
$\chi_1$ (eq.~\ref{chi1ext}) up to the constant $\delta J$.  The two
constants are found by imposing that the solutions for $u_1$ and
$\chi_1$ given by eqs. (\ref{gen1a}) and by
eqs. (\ref{newsol1},\ref{chi1ext}) match continuously at $r=R$.

It should be noted that computing  $\delta J$   by this procedure 
or by eq. (\ref{mom2}) is equivalent; indeed we find a
relative difference smaller than $10^{-10}$. 
This is a further  check of the correctness of the  analytic expression (\ref{effe}) 
for $F(r)$ we derived.

%%%%%%%%%%%%%%%%%%%%%%%%%%%%%%%%%%%%%%%%%%%%%%%%%%%%%%%%%%%%%%%%%%%
\subsubsection{The equations for $w_3$}
%%%%%%%%%%%%%%%%%%%%%%%%%%%%%%%%%%%%%%%%%%%%%%%%%%%%%%%%%%%%%%%%%%%
The solution for the remaining function $w_3$ can be found along the same path.
We start with the equation
\be
\frac{1}{r^4}\frac{d}{dr}(r^4j\frac{dw_3(r)}{dr})+\frac{4}{r}\frac{dj}{dr}w_3
-\frac{10jw_3}{r(r-2M)}=\frac{1}{5}D_2
\label{eqw3}
\ee
and assume $r \le R$;
we introduce the functions 
\be
\label{defu3chi3}
\chi_3=jw_3\qquad\qquad u_3=r^4j\frac{dw_3}{dr}
\ee
and find
\beq
\label{newu3chi3}
&&{\frac{d\chi_3}{dr}=\frac{u_3}{r^4}-\frac{4\pi
r^2(\epsilon+P)\chi_3}{r-2M}},\\\nonumber
&&{\frac{du_3}{dr}=\frac{16\pi r^5(\epsilon+P)\chi_3}{r-2M}}
+\frac{10\chi_3r^3}{(r-2M)}+\frac{r^4}{5}D_2.
\eeq
The general solution of eqs. (\ref{newu3chi3}) for $r \le R$ is
\beq
\label{gen1b}
&&\chi_3=\chi^P_3 + C~\chi^H_3\\\nonumber
&&u_3=u^P_3 + C~u_3^H.
\eeq 
$C$ is a constant and 
$(\chi^P_3,u_3)$ are a particular solution found
by integrating eqs. (\ref{newu3chi3}) with the initial  conditions 
\[
\chi^P_3 (0) =0,\qquad u^P_3(0) =0,\]
$(\chi^H_3, u_3^H)$ are the solution  of the homogeneous system 
obtained by putting $D_2=0$  in  (\ref{newu3chi3}) and integrating 
up to $R$  with the boundary condition
\be
r \rightarrow 0~\quad
\chi^H_3(r)\sim  r^2, \qquad
u_3^H(r)\sim 2~r^5.
\ee

For $r \ge R$ eqs. (\ref{newu3chi3}) reduce to
\beq
\label{extsol3}
&&{\frac{d\chi_3}{dr}=\frac{u_3}{r^4}},\\\nonumber
&&{\frac{du_3}{dr}=\frac{10\chi_3r^3}{(r-2M)}+\frac{u}{5}
\frac{d}{dr}[4k_2-\frac{6J^2}{r^4}]}.
\eeq
The solution of this system is found as a linear combination of the solution 
$\chi^H_{3_{ext}}, u^H_{3_{ext}}$ of the homogenous system found 
by putting $k_2=J=0$ in  (\ref{extsol3}),
and of a particular solution of  (\ref{extsol3}),
$\chi^P_{3_{ext}}, u^P_{3_{ext}}$
which falls to infinity faster than $r^{-5}$.
Both solutions have been calculated by making use of MAPLE.

The homogeneous system  admits two solutions, which, for $r\rightarrow\infty$, behave as 
$\sim r^{-5}$ and $\sim r^2$.  Since the metric must be asymptotically flat,
we have excluded the latter.  

Thus the general solution for $r \ge R$ is
\beq
\label{extgew3}
&&\chi_{3_{ext}}=D\chi^H_{3_{ext}}+\chi^P_{3_{ext}},\\\nonumber
&&u_{3_{ext}}=Du^H_{3_{ext}}+u^P_{3_{ext}},
\eeq
and consequently
\be
w_{3_{ext}}=Dw^H_{3_{ext}}+w^P_{3_{ext}}
\ee
where
\beq
\label{ext1}
&&w^H_{3_{ext}}(r)=(\frac{7}{64r^3M^7})\left[-120r^3M^2
ln(\frac{r}{r-2M})
\right.\\\nonumber
&&\left. +150r^4Mln(\frac{r}{r-2M})
-45r^5ln(\frac{r}{r-2M})
\right.\\\nonumber
&&\left. +20rM^4+60r^2M^3-210r^3M^2+90r^4M+8M^5\right]
\eeq
and
\beq
\label{ext2}
&&w^P_{3_{ext}}(r)=(\frac{J}{480r^7M^9})
\left[(1152J^2M^9+700r^5J^2M^4
\right.\nn\\
&&\left.
+2100r^6J^2M^3-7350r^7J^2M^2+3150r^8J^2M
\right.\nn\\
&&\left.
+280r^4J^2M^5+ 1152Kr^3M^{10}+64J^2rM^8-320J^2r^2M^7
\right.\nn\\
&&\left.+10080KM^5r^8+1088KM^8r^5+1664KM^9r^4\right.\nn\\
&&\left.-23520KM^6r^7+6720KM^7r^6+
16800KM^5r^8ln(\frac{r}{r-2M})\right.\nn\\
&&\left.-4200r^7J^2M^2ln(\frac{r}{r-2M})
+5250r^8J^2Mln(\frac{r}{r-2M})\right.\nn\\
&&\left.-5040KM^4r^9ln(\frac{r}{r-2M})-
13440KM^6r^7ln(\frac{r}{r-2M})\right.\nn\\
&&\left.+576KM^7r^6ln(\frac{r}{r-2M})-
960KM^8r^5ln(\frac{r}{r-2M})\right.\nn\\
&&\left.-384KM^9r^4ln(\frac{r}{r-2M})-1575r^9J^2ln(\frac{r}{r-2M})\right].
\eeq
The constants $C$ and $D$ are found by matching the interior and the exterior solutions at
$r=R$.

%%%%%%%%%%%%%%%%%%%%%%%%%%%%%%%%%%%%%%%%%%%%%%%%%%%%%%%%%%%%%%%%%%%%%%%%%%%%%
%%%%%%%%%%%%%%%%%%%%%%%%%%%%%%%%%%%%%%%%%%%%%%%%%%%%%%%%%%%%%%%%%%%%%%%%%%%%%

\label{lastpage}
\end{document}